\documentclass{JFM-FLM_Au} 

\usepackage{amsmath}
\usepackage{booktabs}
\usepackage{amssymb}
\usepackage{dirtytalk} 
\usepackage{xcolor}
\usepackage{siunitx}

\usepackage{tabularx}
\usepackage{graphicx}
\usepackage{epstopdf}
\usepackage{overpic}
\usepackage{subcaption}
\usepackage{hyperref}

\hypersetup{linkcolor=black}

\lefttitle{Kiran et al.}
\righttitle{Journal of Fluid Mechanics}

\title{The Structure of Merging Turbulent Jets Beneath a Small Quadrotor}

\author{Anoop Kiran\aff{1}\corresp{Email address for correspondence: \href{mailto: anoop_kiran@brown.edu}{anoop\_kiran@brown.edu}}, Nora Ayanian\aff{1, 2} \and Kenneth Breuer\aff{1}}

\affiliation{\aff{1}Center for Fluid Mechanics, School of Engineering, Brown University, Providence, Rhode Island, USA
{\aff{2}Department of Computer Science, Brown University, Providence, Rhode Island, USA}}

\corresau{}

\makeatletter
\gdef\@corresau{}
\def\ps@titlepage{\leftskip\z@\let\@mkboth\@gobbletwo\vfuzz=5\p@
  \def\@oddhead{}\def\@evenhead{}%
  \def\@oddfoot{\hfil\thepage\hfil}\def\@evenfoot{\hfil\thepage\hfil}%
  \def\sectionmark##1{}\def\subsectionmark##1{}}
\makeatother

\begin{document}
\pagestyle{plain} 
\maketitle

\begin{abstract}
The downwash wake of a hovering quadrotor governs both the vehicle's own performance and the safe spacing of multi-rotor formations. Prior measurements have largely characterized the mean flow, using single-point anemometry, volumetric tracking, or planar cuts through part of the rotor system. Higher-order turbulent statistics of the merged wake, and how they relate to canonical jet scaling, have remained unresolved, particularly for small quadrotors at the low-Reynolds-number end of the size range. Here, we present a detailed particle image velocimetry (PIV) study of the downwash of a hovering Crazyflie 2.1 quadrotor (arm length, $l = 46$ mm), sampled along a diagonal cut, passing through rotors along the symmetry axis of the quadrotor, and a front-rotor cut, passing through adjacent rotors. The four rotor jets merge into a single column by $z/l \approx 5$, beyond which the mean velocity profiles progressively approach the canonical round-jet self-similar form, collapsing by $z/l \approx 13$ when scaled by the local centerline velocity and half-width. Centerline decay and half-width growth follow canonical scaling laws with an effective source diameter $D_\text{eff} = 2.29\,l$, effective Reynolds number $\Rey_{D_\text{eff}} = 3 \times 10^4$, at the low end of the range over which canonical jet scaling has been established, and spreading and decay constants nonetheless within the canonical round-jet range. Resolving both cuts shows that the turbulent normal stresses retain a bimodal, cut-dependent signature of the four-rotor source throughout the measurement domain.
\end{abstract}


\section{Introduction}

Quadrotors have improved dramatically in their capabilities, and have been widely deployed in more and more applications including sensing \citep{NakataSurfaceSensing, ErcolaniGasSensing}, atmospheric boundary layer characterization \citep{ThroneberryWakeSampling, LoubimovBoundaryUAV}, and cooperative multi-robot operations \citep{ProrokCollabLocalization, HonigTrajPlanning, KumarLocalizationMapping}. Their rotors generate thrust by accelerating the surrounding air into an axial jet, referred to as~\emph{downwash}, that extends well below the vehicle. The downwash of a leading vehicle can impinge on a nearby vehicle when multiple quadrotors operate in close formation, reducing control authority and leading to loss of thrust and instability \citep{PreissDownwashPlanning, JainDownwashModel, KiranDownwashSeparation}. Establishing safe separation thresholds, therefore, requires a quantitative characterization of both the spatial extent and the turbulence structure of the downwash velocity field.

Studies of quadrotor downwash have focused almost exclusively on the mean flow. Prior work has resolved near-field rotor-rotor interactions and ground effects through extensive studies~\citep{YoonComputationalAerodynamicModeling, ShuklaMultirotorInteractions, LeeRotorInteraction, ThroneberryWakeReview, CarterInfluenceGroundCeiling}. Experimental work has used hot-wire anemometers, smoke visualization, or volumetric PIV to characterize specific configurations: forward and vertical flight in a wind tunnel \citep{ThroneberryAscendFlight}, free-flying quadrotors using Shake-the-Box tracking~\citep{WolfQuadShakethebox}, ground effect on micro-vehicles~\citep{KanGroundEffect, CarterInfluenceGroundCeiling}, and water-to-air transitions for amphibious rotorcraft \citep{HontsAmphibiousPropeller}. \citet{BauersfeldRoboticsMeetsFluidDynamics} provided a detailed mean-flow characterization using a single-point hot-ball anemometer to sample the downwash of six hovering quadrotors ranging from 230~\si{g} to 6.3~\si{kg}. Their results showed that the mean centerline velocity decays with downstream distance in a manner similar to that of a canonical turbulent jet. More recently,~\citet{CeneShearInflow} applied sensor tracking velocimetry to a free-flying DJI NEO in hover and inflow conditions, capturing time-averaged velocity and pressure fields on a reconstruction grid that show the rotor jet merging and the flow's response to uniform and shear inflows.~\citet{KiranInfluenceStaticDynamic} performed planar PIV on quadrotor downwash, but limited their measurement to the plane intersecting the adjacent two of the four rotors. In a related simplified configuration,~\citet{ZhouRotorRotor} used planar and stereoscopic PIV on a side-by-side twin-rotor rig and found the mean thrust and time-averaged flow field to be nearly independent of rotor spacing, while the turbulent kinetic energy and thrust fluctuations rose sharply as the rotors were brought close together. Consequently, a detailed characterization of the merged jet scaling laws, including effective source diameter, spreading rates, and centerline decay constants, and their comparison with canonical free shear flows remains absent.

The turbulent round jet is among the most thoroughly studied flows in fluid mechanics, and its far-field obeys well-established scaling laws. Early measurements~\citep{WygnanskiPreservingJet} established the inverse-linear centerline decay and linear half-width growth, $u_c \propto B D/(z - z_0)$ and $r_{1/2} \propto S(z - z_0)$, where $B$ is the decay constant, $S$ the spreading rate, and $z_{0}$ the virtual origin. Subsequent experiments refined these scaling constants and characterized the higher-order moments~\citep{PanchapakesanAxisymmetricJet, HusseinRoundJet, FukushimaTurbulentJet}, and direct numerical simulations have provided complete energy-budget information~\citep{BoersmaJetDNS, BogeyRoundJetLES}. Across high-Reynolds-number axisymmetric jets, decades of measurements have converged on $B \approx 5.7$--$6.1$, $S \approx 0.086$--$0.096$, and $|z_0|/D \approx 3$--$4$~\citep{WygnanskiPreservingJet, PanchapakesanAxisymmetricJet, HusseinRoundJet}. The merged downwash of a hovering quadrotor, in which four discrete rotor jets coalesce into a single axial column, bears a quantitative resemblance to this canonical flow, though it remains an open question whether that resemblance holds for small quadrotors at the lower end of the size spectrum, where low-Reynolds-number effects become significant~\citep{WinslowLowRe}.

\subsection{Merging jets}
When two or more jets discharge in parallel and merge downstream, the mean flow approaches single-jet self-similarity relatively quickly, but the Reynolds stresses retain a memory of the original multi-source structure for a substantial distance beyond the geometric merging point. This behavior was first documented for dual planar jets~\citep{MillerDualJet, TanakaParallelJets, LinParallelJets}, later extended to twin circular jets \citep{OkamotoTurbulentJet, HarimaTwinJets}, and quantified more recently by PIV and LES \citep{NasrParallelJets, AndersonParallelJets, HarimaTwinJets, LeeParallelJets, LiTwinJets}. In this sense, quadrotor downwash is a four-source variant of the parallel-jet problem, and this body of work provides a direct prediction: the merged jet should exhibit canonical round-jet behavior in its mean field while retaining a four-rotor signature in its second-order statistics.

Compounding this problem is the spatial complexity of the transition itself. Near the rotor plane, the four-source geometry dominates the flow, producing localized structures that depend on the sampled azimuthal direction before eventually recovering a single, quasi-axisymmetric state far downstream. Resolving this transition from a structured, multi-source near-field to an axisymmetric far-field requires an experimental approach capable of mapping both the continuous streamwise evolution and the structural symmetry of the merged jet. 

To bridge this gap, the present study provides detailed PIV-based measurements of the downwash of a small quadrotor that fully resolves both the mean flow and the second-order turbulent statistics along two orthogonal cutting planes through the wake. Sampling the wake on more than one plane is what makes the merging visible as a function of source orientation, allowing us to characterize how the four jets merge and to assess the relaxation of this complex flow toward a single jet governed by canonical scaling laws.

\section{Methodology}
\label{sec:methodology}
The quadrotor used in this study is the Bitcraze Crazyflie 2.1, documented by \citet{HonigFlightROS}, which has a mass $m = 32\,\si{g}$, motor-arm length $l = 46\,\si{mm}$, and rotor diameter $D = 45\,\si{mm}$. 

Velocities are normalized by the rotor-induced velocity from actuator-disk (momentum) theory~\citep{LeishmanHeli},
\begin{equation}
    U_i = \sqrt{\frac{T}{2 \rho \pi R^2}} = \sqrt{\frac{m g}{8 \rho \pi R^2}},
    \label{eq:Ui}
\end{equation}
where $T = mg/4$ is the per-rotor thrust at hover, $\rho = 1.29\,\si{\kilogram\per\meter\cubed}$ is the air density at room conditions, $R = D/2$ is the rotor radius, and $g$ is the gravitational acceleration. For the Crazyflie 2.1, the induced velocity $U_i \approx 4.36\,\si{m/s}$, serving as the reference velocity.

The vehicle was held in a steady hover throughout each measurement campaign using a tether attachment on an $x-z$ translation stage, similar to that used by \citet{KiranDownwashSeparation}. Velocity fields above and below the rotors were quantified using particle image velocimetry (PIV), and to capture the far-field wake of the single quadrotor over several downstream distances spanning the merged wake, the quadrotor was traversed vertically while keeping the laser sheet and camera stationary. This allowed the flow field to be imaged in discrete sections, whose averaged velocity fields were subsequently merged into a composite stitched image spanning a larger vertical extent as in prior work~\cite{KiranDownwashSeparation}. Five sections were acquired at successive traverse positions, covering a streamwise range of approximately $-1.5 \lesssim z/l \lesssim 17.5$ relative to the rotor plane.

\begin{figure}
  \centering
  \begin{subfigure}[b]{0.48\linewidth}
    \centering
    \includegraphics[width=\linewidth]{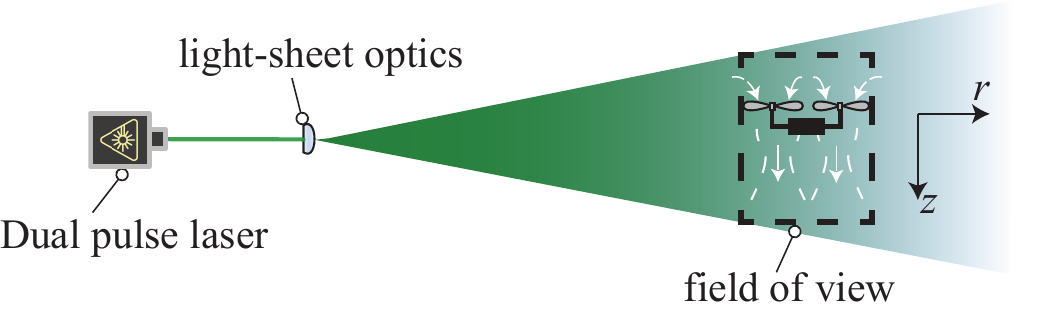}
    \caption{Side view}
    \label{fig:experimental_setup_side}
  \end{subfigure}
  \hfill
  \begin{subfigure}[b]{0.48\linewidth}
    \centering
    \includegraphics[width=\linewidth]{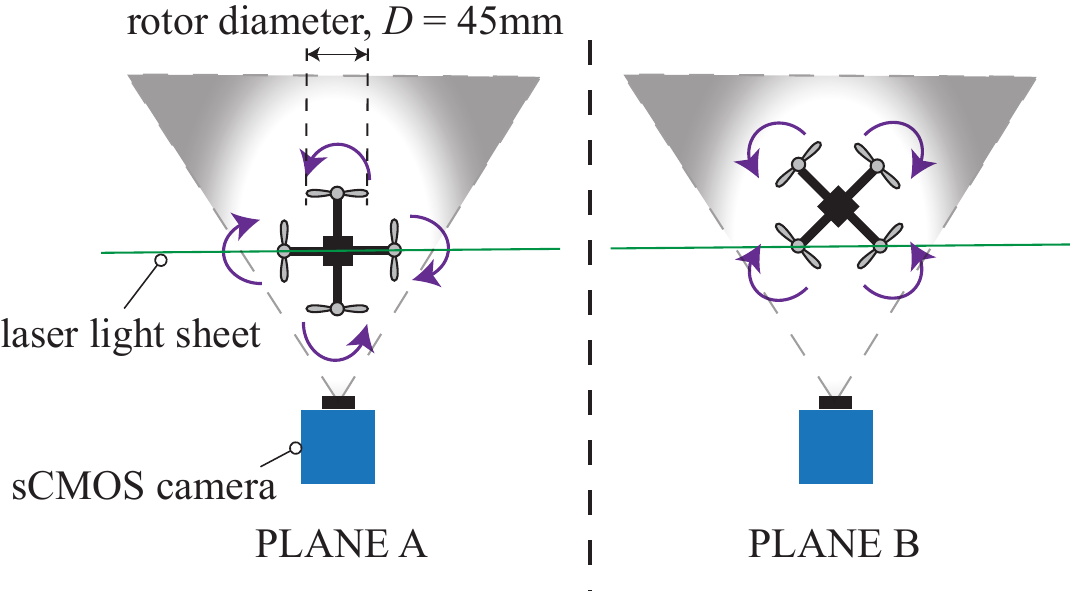}
    \caption{Top view}
    \label{fig:experimental_setup_top}
  \end{subfigure}
  \vspace{1em}
  \hfill
  \begin{subfigure}[b]{0.35\linewidth}
    \centering
    \includegraphics[width=\linewidth]{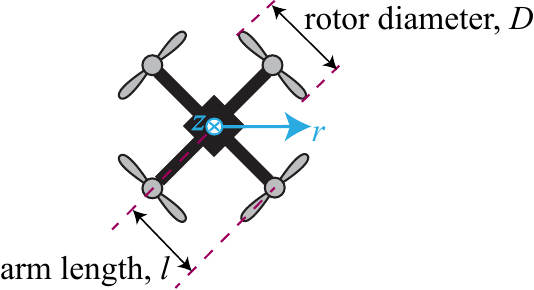}
    \caption{Airframe geometry}
    \label{fig:experimental_setup_geom}
  \end{subfigure}
  \hfill
  \begin{subfigure}[b]{0.48\linewidth}
    \centering
    \includegraphics[width=\linewidth]{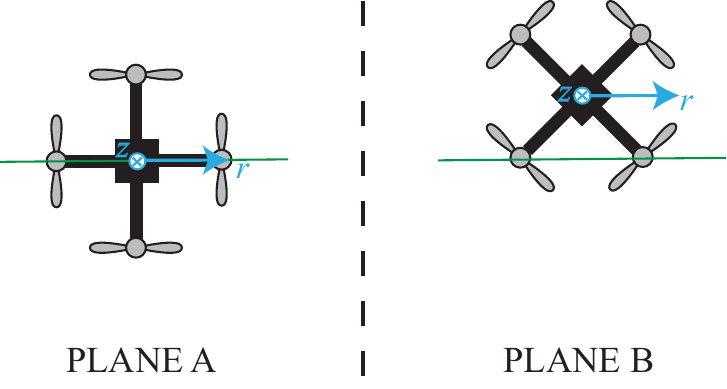}
    \caption{Coordinate system and geometry}
    \label{fig:experimental_setup_coords}
  \end{subfigure}
  \caption{Experimental setup and coordinate frames for the planar two-component PIV measurements of the hovering quadrotor downwash. (\textit{a}) Side-view schematic of the optical arrangement, laser sheet, and field of view (dashed box). The coordinate frame ($r, z$) originates at the airframe center, with $z$ streamwise (positive downward) and $r$ cross-stream. (\textit{b}) Top-view of the two measurement planes, the diagonal cut (plane A) and the front-rotor cut (plane B); curved arrows indicate counter-rotating adjacent rotors. (\textit{c}) Airframe geometry defining the arm length $l$ (central axis to rotor hub) and rotor diameter $D$. (\textit{d}) Cross-stream coordinate $r$ for the two planes, which share an origin but span the diagonal chord in plane A and the shorter inter-rotor chord in plane B.}
  \label{fig:experimental_setup}
\end{figure}

The test volume was seeded with droplets of Di-Ethyl-Hexyl-Sebacic-Acid-Ester (DEHS) tracer particles (${\cal{O}}(1) \mu$m) to track the flow induced by the rotor downwash of the quadrotor. Illumination was provided by a Quantel Evergreen dual-cavity Nd:YAG laser operating at $532\,\si{nm}$ with a maximum pulse energy of $200\,\si{mJ}$ at $15\,\si{Hz}$ as shown by the setup in figure~\ref{fig:experimental_setup_side}. Planar PIV measurements were acquired along two perpendicular cut planes through the wake as indicated in figure~\ref{fig:experimental_setup_top}. The diagonal cut (Plane A) passes through the center of the quadrotor structure, intersecting two opposing rotors at the maximum diagonal distance from the body center. In this plane, the two intersected rotors are located at $r/l = \pm 1$, where $l$ is the motor-arm length as shown in figure~\ref{fig:experimental_setup_geom}. The front-rotor cut (Plane B) passes through two adjacent rotors; in this plane, the two intersected rotors are located at $r/l = \pm 1/\sqrt{2}$. Note that we use the same notation for the radial coordinate, $r$, to denote the distance from the symmetry axis defined by each of the two cutting planes. 

Image sequences were acquired with a LaVision Imager sCMOS camera ($2560 \times 2160$ px) fitted with a Nikon $35\,\si{mm}$ lens, yielding a $520 \times 430\,\si{mm}$ field of view. The acquisition consisted of 1000 image pairs collected over a $50\,\si{s}$ window. Post-processing was carried out using DaVis v10 (LaVision) via a multi-pass cross-correlation scheme, with the initial pass using a $64 \times 64$ px interrogation window with $50\%$ overlap, followed by a final pass using a $32 \times 32$ px interrogation window with $75\%$ overlap. 

Adjacent axial sections were acquired with an overlap of $150\,\si{mm}$, corresponding to roughly a third of the axial ($z$-direction) measurement range. The sections were registered by cross-correlating the velocity fields in the overlap region and combined using a linear feathering interpolation that smoothly weighted the upstream and downstream sections across the boundary. Subsequent post-processing and analyses were performed in $\textsc{MATLAB}$. 

Throughout the paper, the coordinate system is defined with $z$ as the axial (streamwise) coordinate (positive downward, away from the rotor face, parallel to gravity) and $r$ the radial coordinate (cross-stream). The streamwise and cross-stream velocity components are denoted by $u$ and $v$, respectively. Following the standard Reynolds decomposition, the streamwise and cross-stream velocities are written as $u = \langle u \rangle + u'$ and $v = \langle v \rangle + v'$, where $\langle u \rangle$ and $\langle v \rangle$ are the time-averaged means and $u'$ and $v'$ are the fluctuations about those means. The Reynolds normal stresses are $\langle u'u' \rangle$, $\langle v'v' \rangle$, and the Reynolds shear stress is $\langle u'v' \rangle$. 

\begin{figure}
  \centering
  
  \begin{subfigure}[b]{0.48\linewidth}
    \centering
    \begin{overpic}[width=\linewidth]{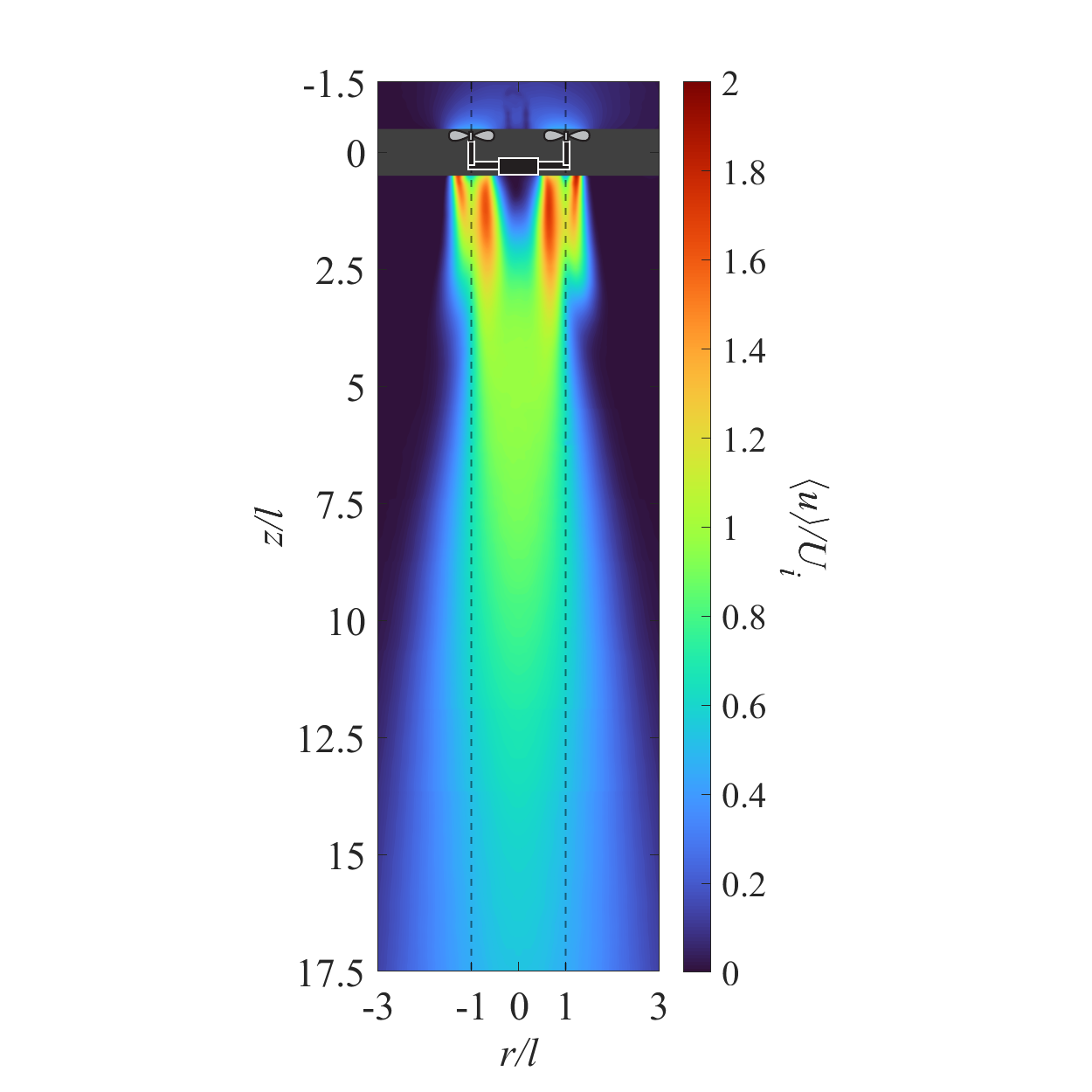}
    \put(75, 75){\includegraphics[width=0.25\linewidth]{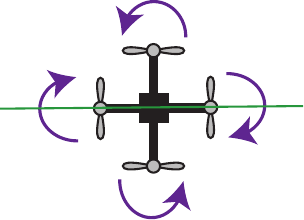}}
    \end{overpic}
    \caption{Diagonal cut, $\langle u\rangle$ contour}
    \label{fig:u_mean_contour_diag}
  \end{subfigure}
  \hfill 
  \begin{subfigure}[b]{0.48\linewidth}
    \centering    
    \begin{overpic}[width=\linewidth]{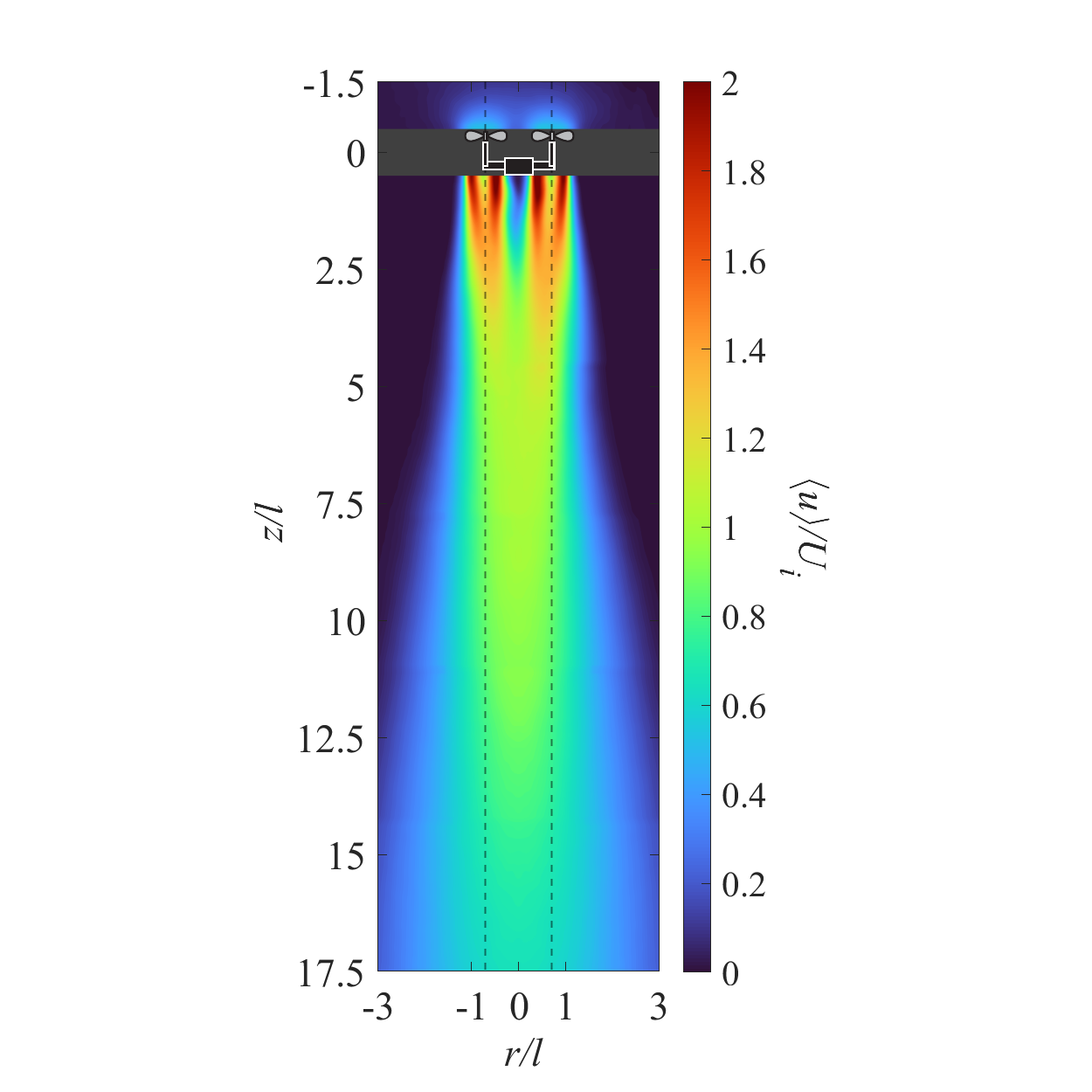}
    \put(75, 78.5){\includegraphics[width=0.25\linewidth]{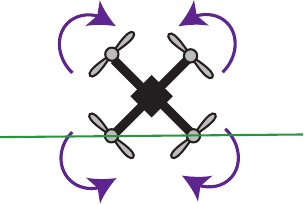}}
    \end{overpic}
    \caption{Front-rotor cut, $\langle u\rangle$ contour}
    \label{fig:u_mean_contour_front}
  \end{subfigure}

  \vspace{6pt} 
  \begin{subfigure}[b]{0.48\linewidth}
    \centering
    \includegraphics[width=\linewidth]{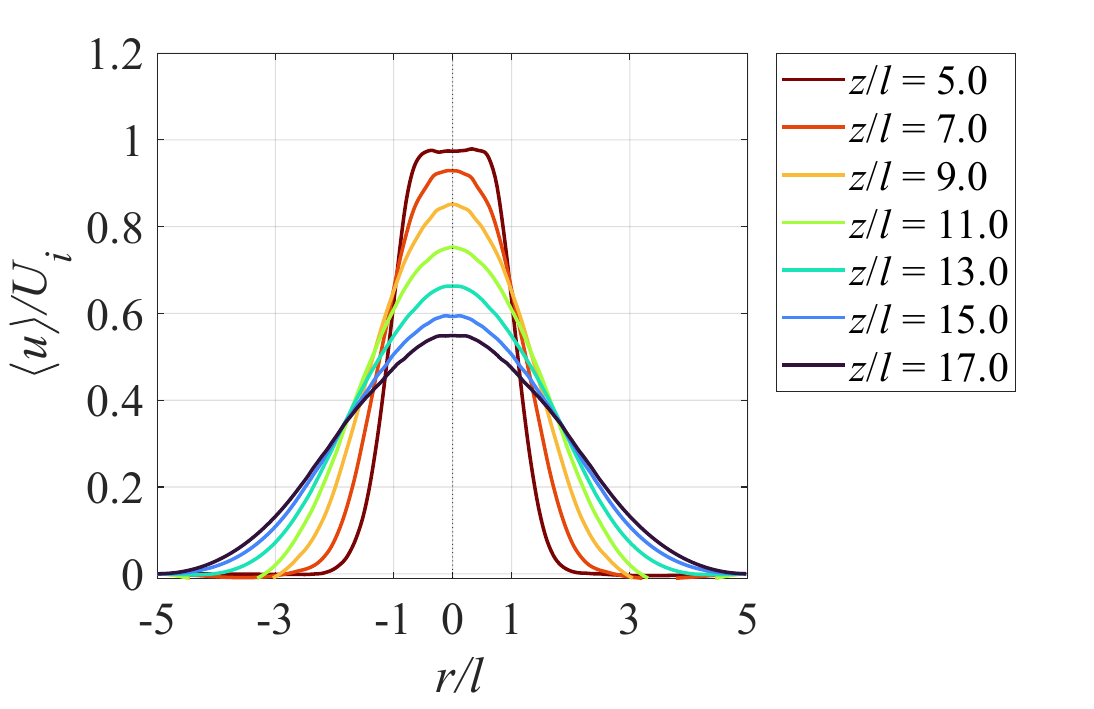}
    \caption{Diagonal cut, $\langle u\rangle$ profiles}
    \label{fig:u_mean_profile_diag}
  \end{subfigure}
  \hfill
  \begin{subfigure}[b]{0.48\linewidth}
    \centering
    \includegraphics[width=\linewidth]{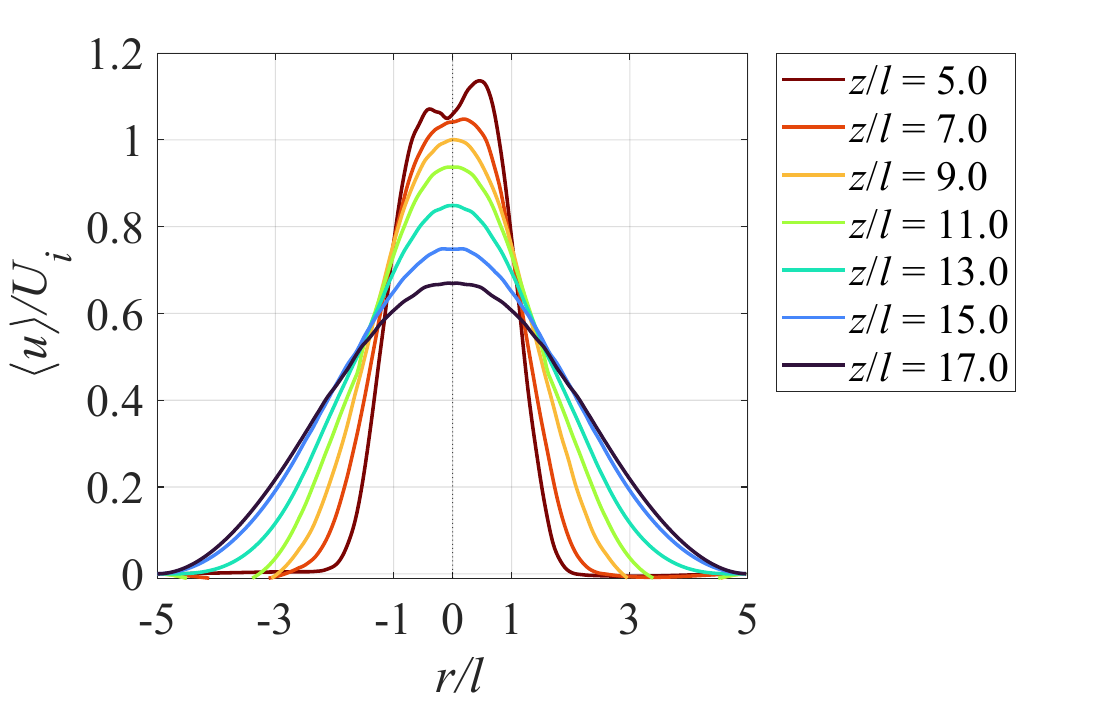}
    \caption{Front-rotor cut, $\langle u\rangle$ profiles}
    \label{fig:u_mean_profile_front}
  \end{subfigure}

  \caption{Mean streamwise velocity $\langle u\rangle/U_i$ of hovering quadrotor downwash measured by planar PIV. Contour panels: (\textit{a}) the diagonal cut through two opposing rotors and (\textit{b}) the front-rotor cut through two front rotors. The gray patch masks the quadrotor body, and an orientation icon in each panel indicates the measurement plane. Dashed vertical lines mark the in-plane rotor positions: $r/l = \pm 1$ in the diagonal cut (\textit{a}) and $r/l = \pm 1/\sqrt{2}$ in the front-rotor cut (\textit{b}). Profile panels: (\textit{c}, \textit{d}) cross-stream profiles at downstream stations $z/l \in \{5, 7, 9, 11, 13, 15, 17\}$, cooler colors farther downstream. Velocities are normalized by the induced velocity $U_i$~\eqref{eq:Ui}; profiles beyond $|r|/l = 3$ are extrapolated for visualization.}
  \label{fig:u_velocity_mean}
\end{figure}

\section{Results and Discussion}
\subsection{Mean flow field}
\label{subsec:mean_flow}

\subsubsection{Mean streamwise velocity}
\label{subsubsec:u_mean}

The mean streamwise velocity field, $\langle u \rangle/U_i$, exhibits qualitatively different near-field structures in the two cut planes A \& B (figure~\ref{fig:experimental_setup_top}). The rotor-driven jet structure is apparent immediately below the rotor plane, showing the downward velocities on either side of each of the two rotor axes of rotation. Those flows quickly merge, and by $z/l \approx 2$ two distinct jets appear in each cut, corresponding to the two rotors, separated by a dead zone - a low-velocity core characterized by minimal streamwise momentum. The presence of this dead zone is consistent with prior experiments (\cite{WolfQuadShakethebox, BauersfeldRoboticsMeetsFluidDynamics}).

In the diagonal cut (figure~\ref{fig:u_mean_contour_diag}), the two jets merge early. Its inner core is fuller in the near-field because the two off-plane rotors sit closer to this plane and contribute more strongly to the flow below it, filling the centerline with merged flow from upstream of the cut. The streamwise profile is therefore already nearly single-peaked by $z/l \approx 5$ (figure~\ref{fig:u_mean_profile_diag}) and decays monotonically into the far-field.

The front-cut (figure~\ref{fig:u_mean_contour_front}) merges later, with the two jets remaining distinct well downstream. The bimodal profile persists to $z/l \approx 5$, separated by a centerline depression, before the depression fills in and the profile becomes single-peaked by $z/l \approx 7$ (figure~\ref{fig:u_mean_profile_front}). The merged column, as in the diagonal cut, decays monotonically into the far-field. This later merge follows from the geometry, where the off-plane rotors lie farther from this plane than in the diagonal cut, so the near-field centerline is fed mainly by the two in-plane jets, which also leaves the diagonal-cut centerline velocity consistently lower at a given $z/l$.

\subsubsection{Mean cross-stream velocity}
\label{subsubsec:v_mean}

\begin{figure}[htbp]
  \centering
  
  \begin{subfigure}[b]{0.48\linewidth}
    \centering
    \begin{overpic}[width=\linewidth]{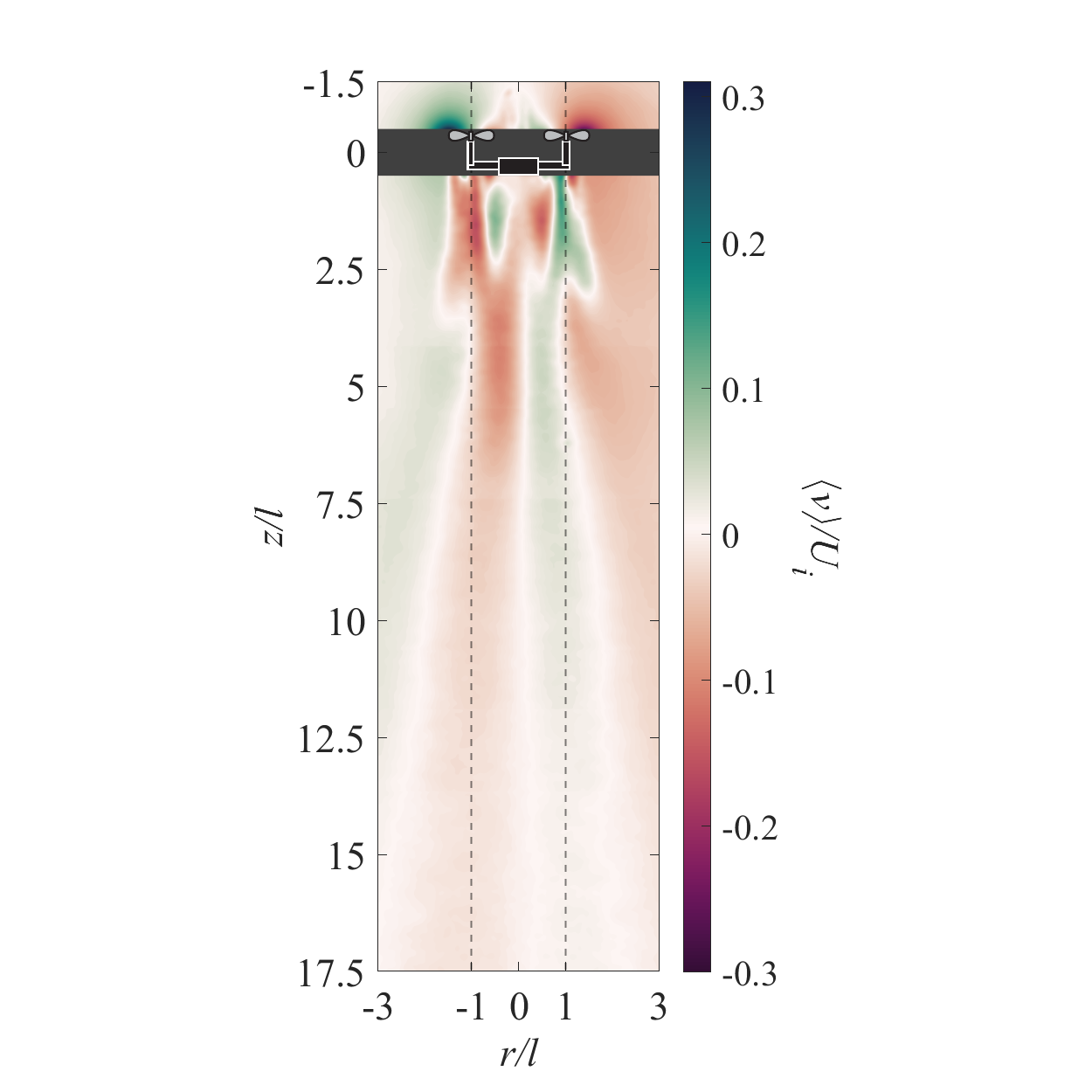}
    \put(75, 75){\includegraphics[width=0.25\linewidth]{figures/PDF_files/diagonal_cut_icon.pdf}}
    \end{overpic}
    \caption{Diagonal cut, $\langle v\rangle$ contour}
    \label{fig:v_mean_contour_diag}
  \end{subfigure}
  \hfill
  \begin{subfigure}[b]{0.48\linewidth}
    \centering
    \begin{overpic}[width=\linewidth]{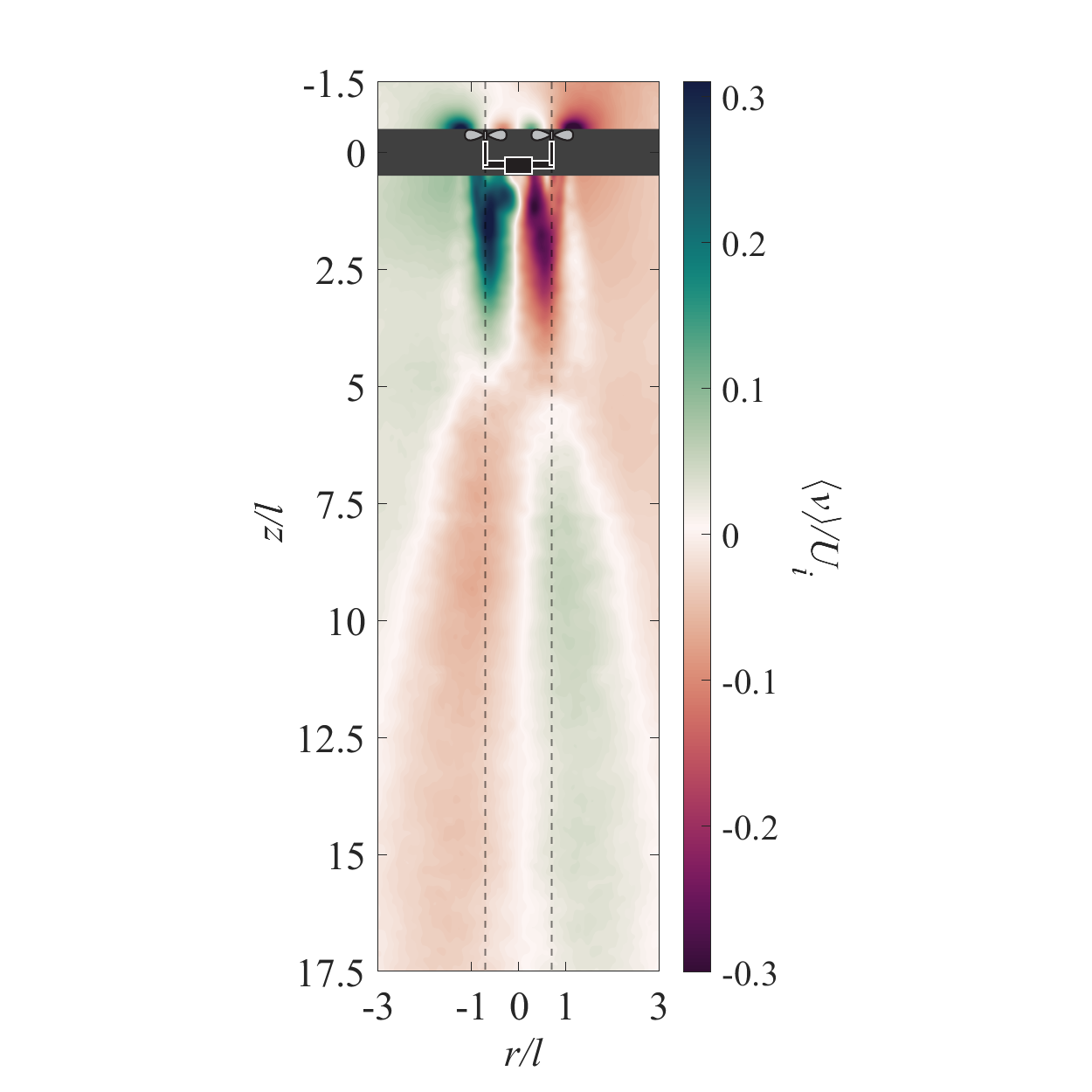}
    \put(75, 78.5){\includegraphics[width=0.25\linewidth]{figures/PDF_files/front_cut_icon.pdf}}
    \end{overpic}
    \caption{Front-rotor cut, $\langle v\rangle$ contour}
    \label{fig:v_mean_contour_front}
  \end{subfigure}

  \vspace{6pt} 
  \begin{subfigure}[b]{0.48\linewidth}
    \centering
    \includegraphics[width=\linewidth]{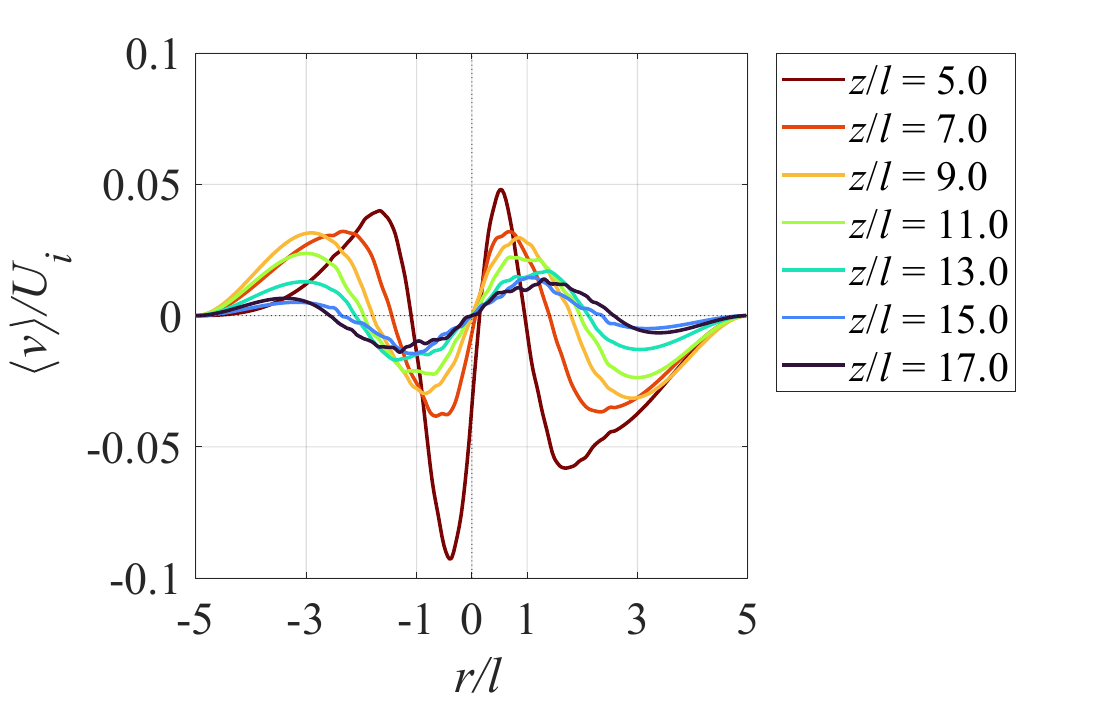}
    \caption{Diagonal cut, $\langle v\rangle$ profiles}
    \label{fig:v_mean_profile_diag}
  \end{subfigure}
  \hfill
  \begin{subfigure}[b]{0.48\linewidth}
    \centering
    \includegraphics[width=\linewidth]{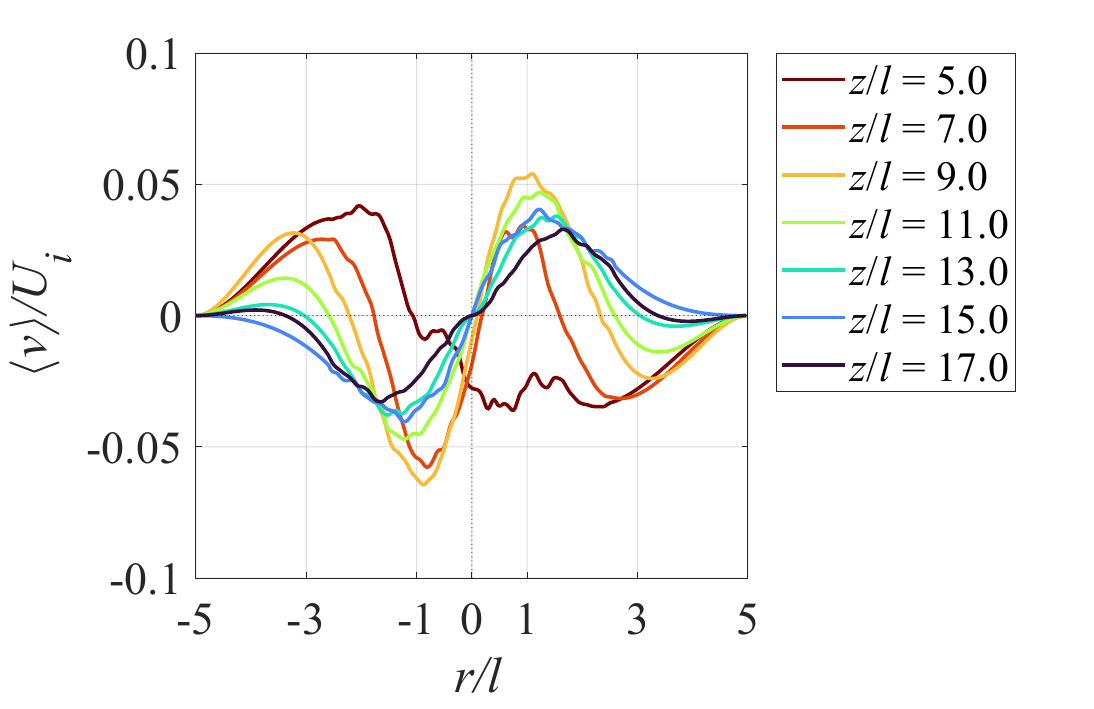}
    \caption{Front-rotor cut, $\langle v\rangle$ profiles}
    \label{fig:v_mean_profile_front}
  \end{subfigure}

  \caption{Mean cross-stream velocity $\langle v\rangle/U_i$ of hovering quadrotor downwash measured by planar PIV. Contour panels: (\textit{a}) the diagonal cut through two opposing rotors and (\textit{b}) the front-rotor cut through two front rotors. The gray patch masks the quadrotor body, and an orientation icon in each panel indicates the measurement plane. Dashed vertical lines mark the in-plane rotor positions: $r/l = \pm 1$ in the diagonal cut (\textit{a}) and $r/l = \pm 1/\sqrt{2}$ in the front-rotor cut (\textit{b}). Profile panels: (\textit{c}, \textit{d}) cross-stream profiles at downstream stations $z/l \in \{5, 7, 9, 11, 13, 15, 17\}$, cooler colors farther downstream. Velocities are normalized by the induced velocity $U_i$~\eqref{eq:Ui}; profiles beyond $|r|/l = 3$ are extrapolated for visualization.}
  \label{fig:v_velocity_mean}
\end{figure}

The mean cross-stream velocity field, $\langle v \rangle/U_i$, exhibits two key patterns: a divergent near-field dominated by the spreading of the individual rotor wakes, and an antisymmetric entrainment pattern characteristic of a turbulent jet as it transitions into the far-field (figures~\ref{fig:v_mean_contour_diag} and~\ref{fig:v_mean_contour_front}). In both cuts, outward-directed lobes appear immediately below each intersected rotor, reflecting the lateral spreading of each jet as it leaves the disk. 

The diagonal-cut (figure~\ref{fig:v_mean_contour_diag}) shows a more complex near-field, where the in-plane jets and the contribution of the two off-plane rotors overlap below the imaged plane. The lobes here are comparatively weak ($|\langle v \rangle|/U_i \approx 0.1$) but persist further downstream before the flow reorganizes. By $z/l \approx 5$ this complex structure has given way to a canonical entrainment pattern, with $\langle v \rangle$ converging radially toward the centerline as ambient fluid is drawn into the wake.

The front-rotor cut (figure~\ref{fig:v_mean_contour_front}) is simpler, with two well-separated lobes of relatively larger magnitude ($|\langle v \rangle|/U_i \approx 0.3$), since the closer in-plane rotor pair spreads more strongly with the imaged plane. The bimodal near-field similarly transitions to radial convergence by $z/l \approx 5$. In both cuts, the transition from divergent to convergent flow spans $2.5 \lesssim z/l \lesssim 5$, where the merging jets expand outward while entrainment draws flow inward, producing stagnation zones in which the lateral flow reverses direction. The peak entrainment magnitude beyond this point is $|\langle v \rangle|/U_i \approx 0.05$ in the front-rotor cut and $\approx 0.03$ in the diagonal cut, dropping to $\lesssim 0.01$ for $|r/l| > 2$.

Across both regions and both cuts, $\langle v \rangle$ vanishes on the centerline ($r/l = 0$): in the near-field the inward contributions from opposing rotor pairs cancel by symmetry, and in the far-field the streamwise velocity peaks on the centerline where the cross-stream component is null.

The two cuts differ because they sample the wake along chords of different length. The stronger in-plane spreading of the closer front-rotor pair broadens the merged column, giving larger half-widths in the front-rotor cut at the same $z/l$ (cf. figures~\ref{fig:u_mean_profile_diag} and~\ref{fig:u_mean_profile_front}) and a slightly elliptical wake cross-section elongated along the front-rotor direction, as reported for merging twin jets~\citep{OkamotoTurbulentJet, TaddesseTwinJetLES}. The diagonal cut, by contrast, lies nearer its two off-plane rotors and so receives a larger streamwise contribution from them, contributing to the centerline-velocity differences discussed in~\S\ref{subsec:centerline_scaling}.

\begin{figure}
  \centering
  \begin{minipage}[c]{0.35\linewidth}
      \centering
      \includegraphics[width=\linewidth]{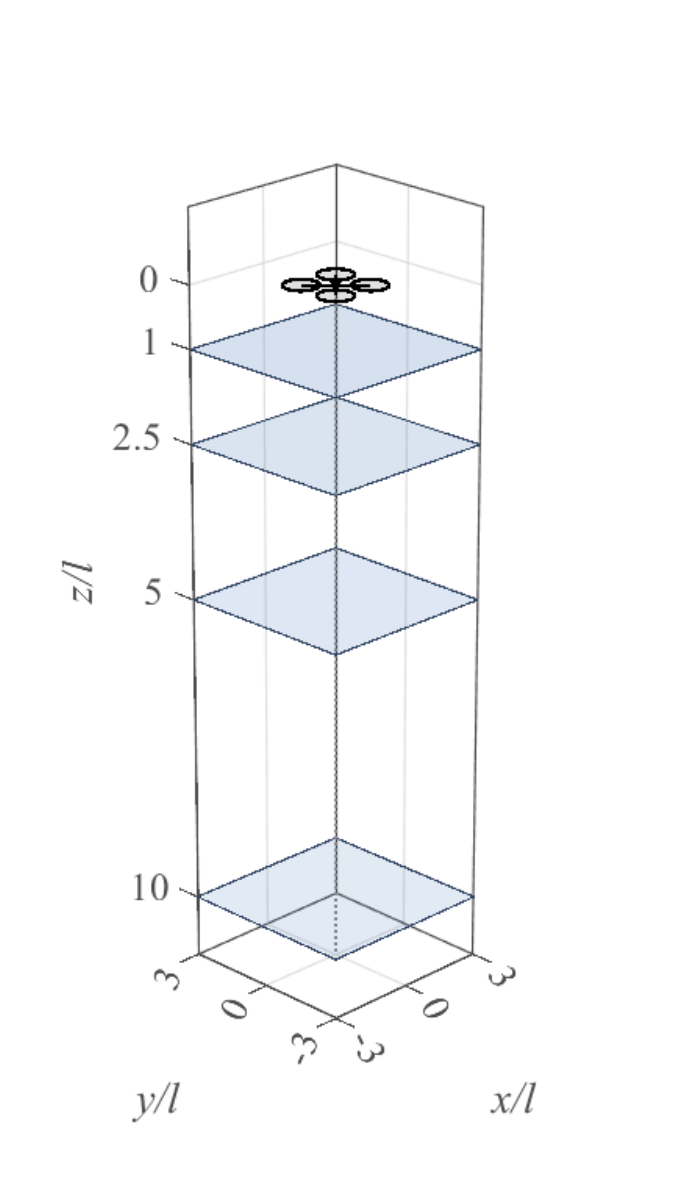}
  \end{minipage}%
  \hfill
  \begin{minipage}[c]{0.65\linewidth}
    \centering
    \begin{subfigure}[b]{0.48\linewidth}
      \centering
      \includegraphics[width=\linewidth]{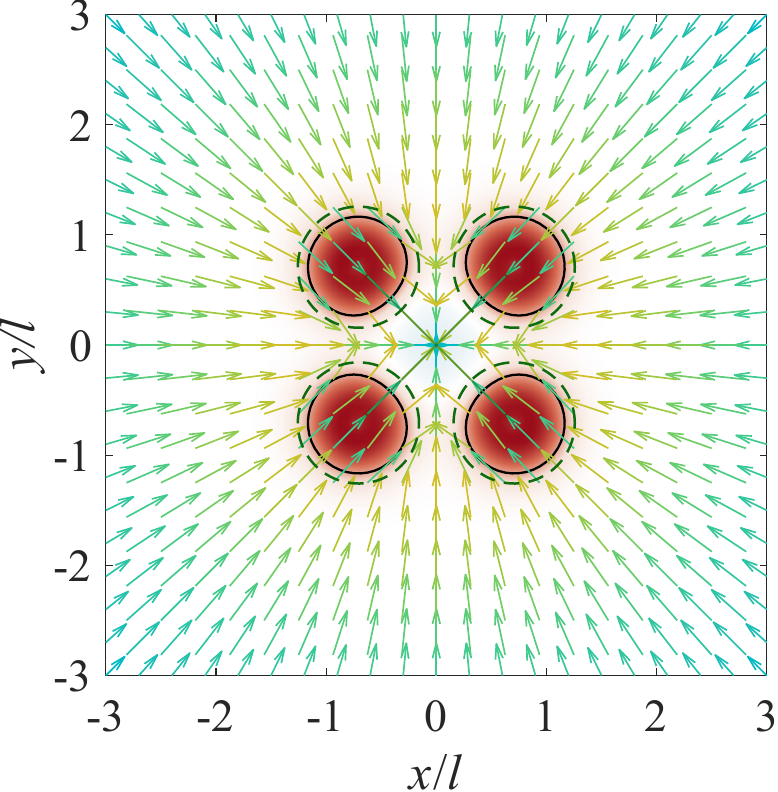}
      \caption{$z/l = 1$}
      \label{fig:unified_schematic_z1}
    \end{subfigure}
    \hfill
    \begin{subfigure}[b]{0.48\linewidth}
      \centering
      \includegraphics[width=\linewidth]{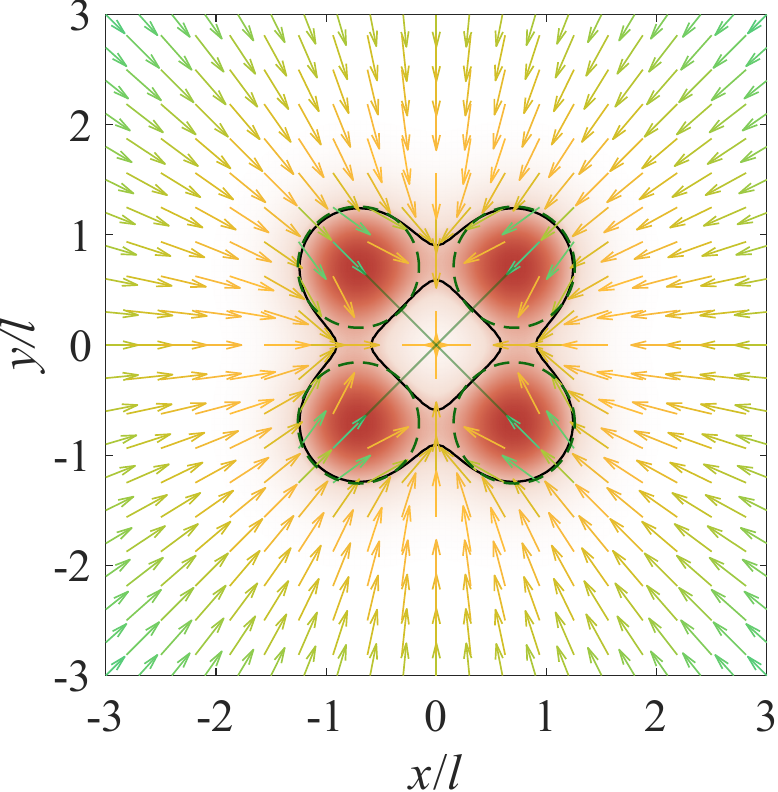}
      \caption{$z/l = 2.5$}
      \label{fig:unified_schematic_z2p5}
    \end{subfigure}

    \vspace{0.6em}

    \begin{subfigure}[b]{0.48\linewidth}
      \centering
      \includegraphics[width=\linewidth]{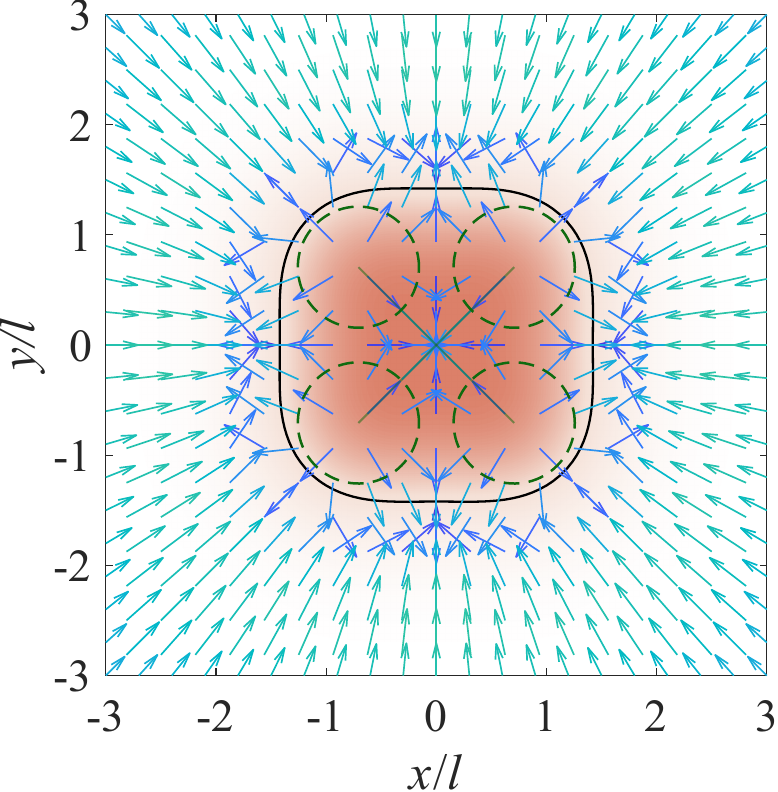}
      \caption{$z/l = 5$}
      \label{fig:unified_schematic_z5}
    \end{subfigure}
    \hfill
    \begin{subfigure}[b]{0.48\linewidth}
      \centering
      \includegraphics[width=\linewidth]{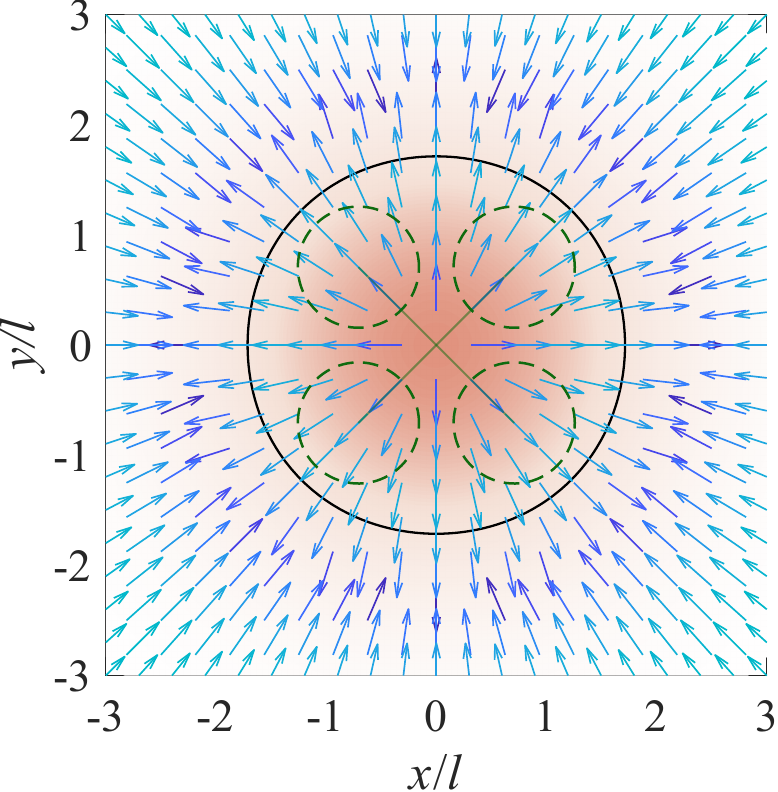}
      \caption{$z/l = 10$}
      \label{fig:unified_schematic_z10}
    \end{subfigure}

    \vspace{0.6em}

    \includegraphics[width=0.98\linewidth, trim=2 2 2 2, clip]{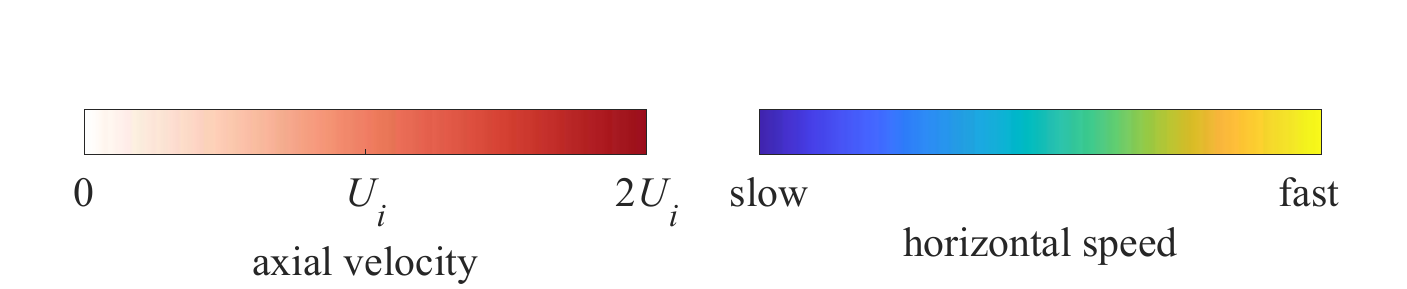}
  \end{minipage}

    \caption{Qualitative top-down reconstruction of the flow beneath the quadrotor at four streamwise stations ($z/l = 1$, $2.5$, $5$, $10$), assembled from the topology across the two orthogonal PIV planes (figures~\ref{fig:u_velocity_mean} and~\ref{fig:v_velocity_mean}). Shaded color shows axial velocity, with downwash (red) strongest just below the disk; the black curve is its half-peak contour, marking the edge of the downwash core. Arrows are drawn at fixed length, and their colors indicate horizontal speed from dark blue/violet (slowest) through cyan and green to yellow (fastest). Dashed green circles mark the rotor disks. The four annular cores at $z/l = 1$ (\textit{a}) merge into a clover-shaped region by $z/l = 2.5$ (\textit{b}), fill into a single column with faint residual squareness by $z/l = 5$ (\textit{c}), and form a near-axisymmetric column of effective diameter $D_\text{eff} \approx 2.29\,l$ by $z/l = 10$ (\textit{d}).
    \label{fig:unified_schematic}}
\end{figure}

\subsubsection{Unified schematic of mean flow development}
\label{subsubsec:unified_schematic}

To synthesize the three-dimensional mean-flow topology from the two orthogonal PIV planes, figure~\ref{fig:unified_schematic} presents top-down reconstructions at four streamwise levels, tracing how the diagonal and front-rotor planes sample different chords of a single merging process.

At $z/l = 1$, each rotor slipstream still acts as a separate momentum source, spreading outward on its own (figure~\ref{fig:unified_schematic_z1}). The four half-peak contours are disjoint; the space between them carries up-flow \citep{NakataSurfaceSensing}, and the in-plane vectors point inward everywhere as ambient fluid is drawn toward each core. The merging of sources depends on how far apart the rotors sit in a given cut. In the diagonal cut, the in-plane pair is widely separated, at $2l$, so the central core fills only gradually as the off-plane jets drift toward the axis. In the front-rotor cut, the pair sits closer, at $\sqrt{2}\,l$, and their shear layers meet sooner, drawing fluid into the central column earlier and finishing the merge upstream of where the diagonal-cut jets do. This is why the front-rotor cut carries the wider wake downstream.

By $z/l = 2.5$ the four contours have joined into one clover-shaped region (figure~\ref{fig:unified_schematic_z2p5}). An inner contour still marks an unfilled hub, so the wake closes at its edges before its center. The vectors are uniformly inward and brightest of the four stations, as the merging cores draw hardest on the surrounding fluid, making entrainment strongest here.

At $z/l = 5$ the hub has filled through contributions from all four sources, and only a faint squareness survives in the outline (figure~\ref{fig:unified_schematic_z5}). The two cuts read the axis differently. In the diagonal cut, $r = 0$ is the wake centerline, where $\langle v \rangle$ tends to $0$ by axisymmetry. In the front-rotor cut, $r = 0$ lies on the symmetry line between two adjacent rotors, where the streamwise flow is weak, and the opposing lateral flows cancel.

Once merged, for $z/l > 5$, the flow is a single downwash column that spreads outward while still entraining ambient fluid (figure~\ref{fig:unified_schematic_z10}). Its in-plane vectors are darker than at $z/l = 2.5$, indicating that the horizontal motion has decayed, and along either cut the streamwise and cross-stream profiles collapse onto the canonical round-jet self-similar form once scaled by the local centerline velocity and half-width (\S\ref{subsec:self_sim_mean_velocity}).

The four-fold footprint in panels~(figure~\ref{fig:unified_schematic}\textit{a}--\textit{c}) follows from rotor placement, the four momentum sources, and the way they merge. A similar pattern appears in the volumetric Shake-The-Box measurements of~\citet{WolfQuadShakethebox}, who resolved a persistent plus-shaped downwash beneath a free-flying quadrotor. They traced its lateral lobes to the rotor rotation directions, with the spin of opposing rotors ejecting fluid outboard. In comparison to the inward-canted rotors in their experiment, the Crazyflie 2.1 carries conventional, uncanted rotors, so its merging is not tilt-driven. The rotation mechanism still applies, though, since both vehicles use counter-rotating pairs. A rotation-driven plus pattern is therefore distinct in origin and symmetry from the spacing-induced structure shown here.

\subsection{Centerline decay and jet width scaling}
\label{subsec:centerline_scaling}

The far-field scaling of the merged wake is characterized through the streamwise centerline velocity $u_c(z)  \equiv \langle u\rangle(r=0,z)$ and the jet half-width $r_{1/2}(z)$, where the latter defines the radial position at which the streamwise velocity drops to half of its centerline value:
\begin{equation}
    \langle u\rangle\!\left(r_{1/2}(z), z\right) = \frac{1}{2}u_c(z).
    \label{eq:half_width_def}
\end{equation}

Downstream of the merge location $z_\text{merge}$, canonical round-jet scaling laws dictate an inverse-linear decay of the centerline velocity alongside a linear growth of the half-width with downstream distance \citep{PopeTurbulentFlows}:
\begin{equation}
    \frac{u_c(z)}{U_i} = \frac{B\,D_\text{eff}}{z - z_0},
    \label{eq:centerline_decay}
\end{equation}
\begin{equation}
    \frac{r_{1/2}(z)}{l} = S\,\frac{z - z_0}{l},
    \label{eq:halfwidth_growth}
\end{equation}
where $B$ represents centerline decay constant, $S$ denotes the spreading rate, $D_\text{eff}$ is the effective source diameter, and $z_0$ is the virtual origin. 

The merge location is identified directly from the empirical data as the minimum downstream $z$ value where the peak streamwise velocity coincides with the centerline value, marking the transition from near-field bimodal structure to unified merged wake. For the diagonal cut, this definition yields $z_\text{merge}/l = 5.0$. The effective source diameter, taken as twice the wake half-width at the merge plane, is given by:
\begin{equation}
    D_\text{eff} = 2\,r_{1/2}(z_\text{merge}),
    \label{eq:Deff}
\end{equation}
resulting in $D_\text{eff}/l = 2.29$ ($\approx 105$mm). 

Fitting equations~\eqref{eq:centerline_decay} and \eqref{eq:halfwidth_growth} to the diagonal-cut data over the merged region ($z/l \ge 5$) yields a centerline decay constant $B = 6.02$, a spreading rate $S = 0.088$, and a virtual origin $z_0/D_\text{eff} = -3.42$. As summarized in table~\ref{tab:scaling_constants}, all three parameters fall within the classical ranges documented in canonical round-jet literature. Quantitative scaling and decay calculations are performed within the experimental data collection region ($|r/l| \le 3$), ensuring that the derived physical constants remain independent of the extrapolated range shown in profile visualizations (figures 2--6 (\textit{c} and \textit{d})). 

\begin{figure}[htbp]
  \centering
  
  \begin{subfigure}[b]{0.48\linewidth}
    \centering
    \includegraphics[width=\linewidth]{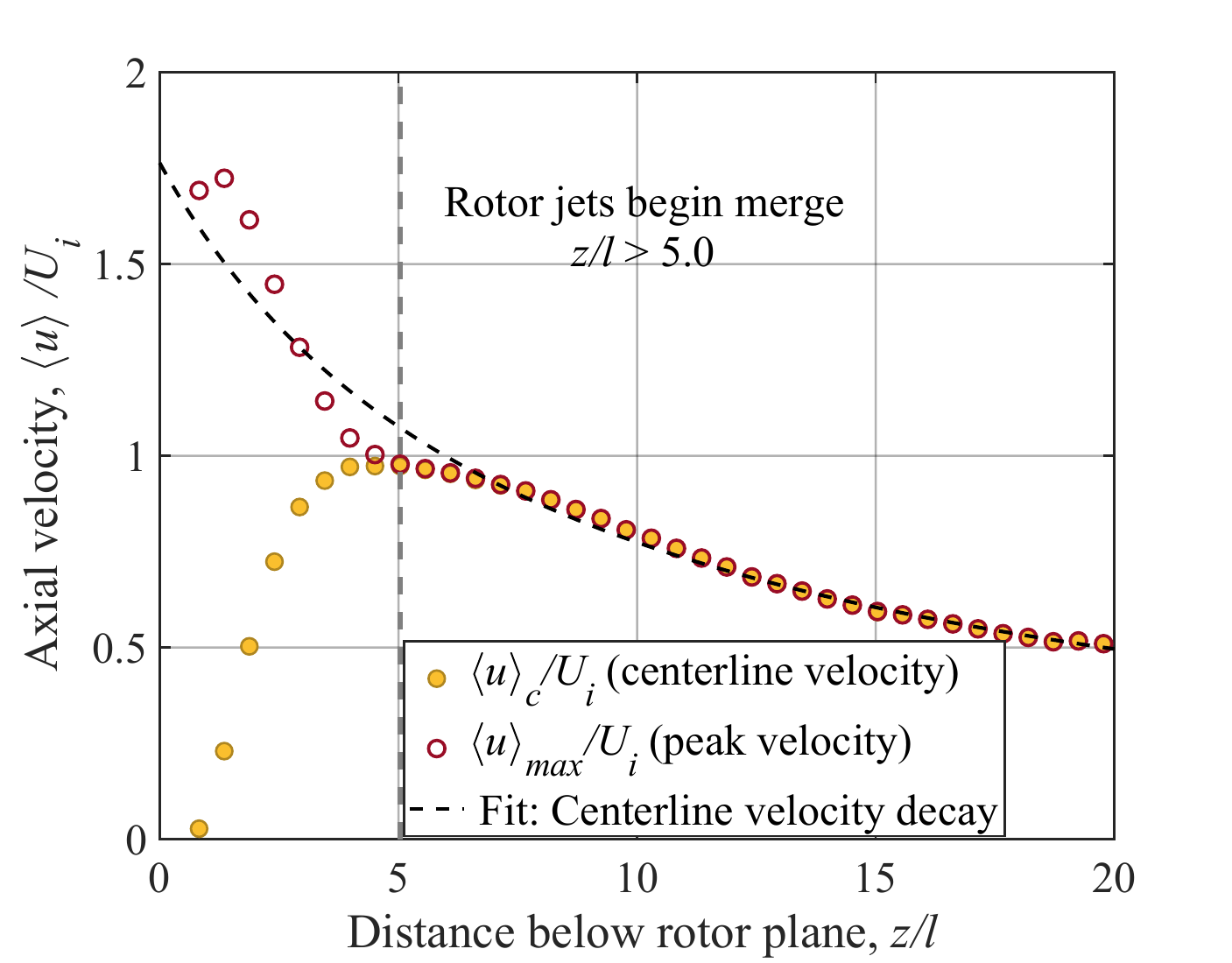}
    \caption{Centerline velocity decay}
    \label{fig:centerline_decay}
  \end{subfigure}
  \hfill
  \begin{subfigure}[b]{0.48\linewidth}
    \centering
    \includegraphics[width=\linewidth]{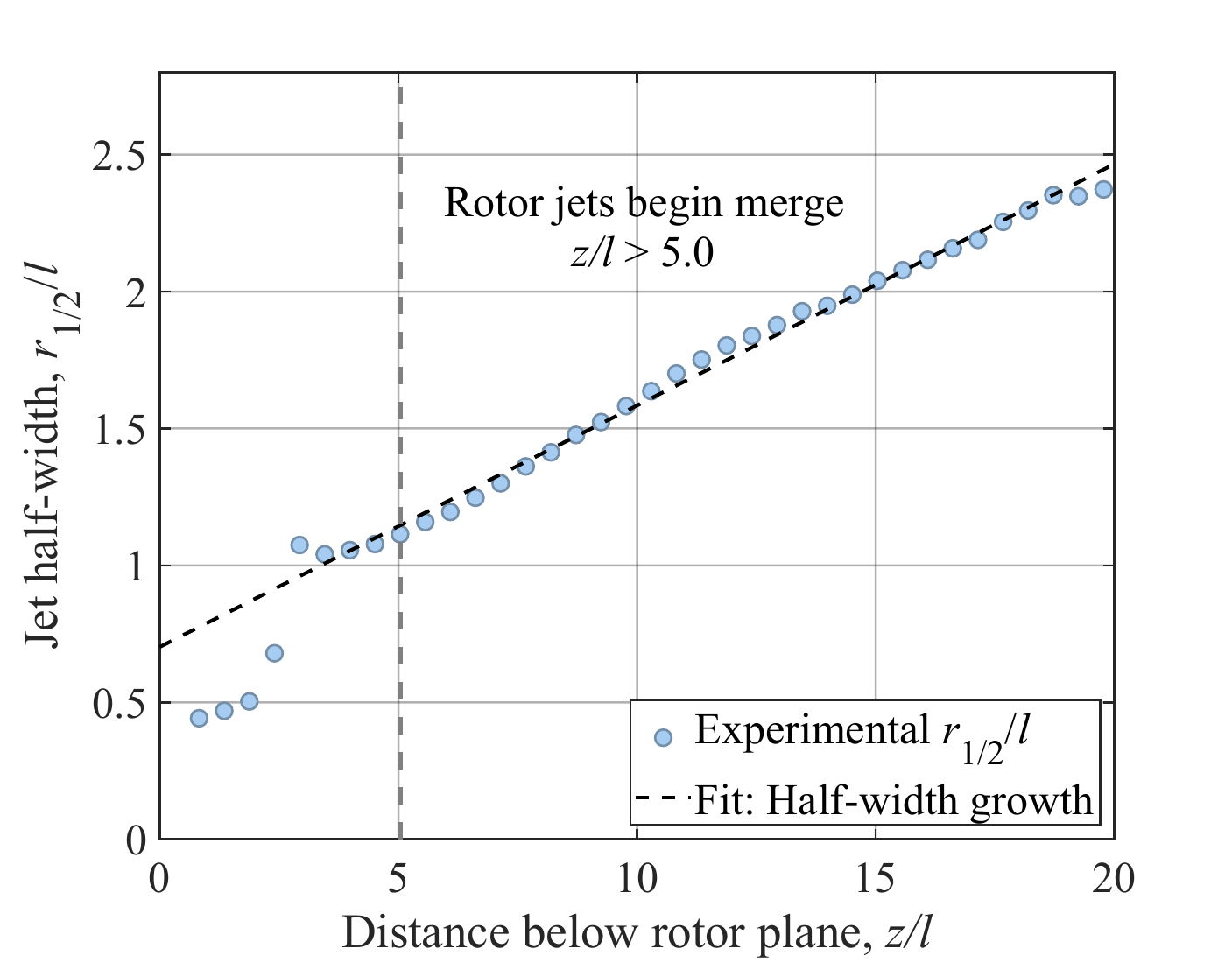}
    \caption{Wake half-width growth}
    \label{fig:width_growth}
  \end{subfigure}

  \caption{Centerline scaling laws of the diagonal-cut wake. (\textit{a}) Centerline streamwise velocity $\langle u \rangle_c/U_i$ (filled circles) and peak in-plane velocity $\langle u \rangle_{\max}/U_i$ (open circles) versus downstream distance $z/l$, with the canonical round-jet hyperbolic fit~\eqref{eq:centerline_decay} (dashed line). (\textit{b}) Wake half-width $r_{1/2}/l$ (filled circles) versus downstream distance $z/l$, with the linear fit~\eqref{eq:halfwidth_growth} (dashed line).} 
  \label{fig:centerline_decay_width_growth}
\end{figure}

\begin{table}
  \begin{center}
  \def~{\hphantom{0}}
  \begin{tabular}{lcccccc}
      \hline
      Reference                               & Method  & Configuration   & $\Rey_{D}$           & $B$   & $S$   & $|z_0|/D$ \\[3pt]
      \hline
      \citet{FukushimaTurbulentJet} & Exp. & Single free jet & $2.0\times10^3$      & 6.70  & 0.096 & 6.8  \\
      \citet{TaubJetDNS} & Sim. & Single free jet & $2.0\times10^3$      & 5.40  & 0.096 & 1.3  \\
      \citet{BoersmaJetDNS}                  & Sim. & Single free jet & $2.4\times10^3$      & 5.90  & 0.093 & 4.9  \\
      \citet{NguyenDNSRoundJet}                    & Sim. & Single free jet & $3.5\times10^3$      & 5.15  & 0.089 & 0.0  \\
      \citet{NguyenRoundJetDNSPOF}                   & Sim. & Single free jet & $7.0\times10^3$      & 5.25  & 0.086 & 2.0  \\
      \citet{LabanTwinJetsExp}                    & Exp. & Twin free jets & $1.0\times10^4$      & 4.76--5.88  & 0.08--0.10 & --  \\
      \citet{PanchapakesanAxisymmetricJet}          & Exp. & Single free jet & $1.1\times10^4$      & 6.06  & 0.096 & 0.0  \\
      \citet{BogeyRoundJetLES}        & Sim. & Single free jet & $1.1\times10^4$      & 6.4  & 0.087 & --  \\
      \citet{TaddesseTwinJetLES}        & Sim. & Single free jet & $1.1\times10^4$      & 6.06  & 0.094 & --  \\
    \citet{BirchFreeJetExp}        & Exp. & Single free jet & $1.6\times10^4$      & 4.0  & 0.097 & 5.8  \\
      \citet{TongTurbulentJet}                  & Exp. & Single free jet & $1.8\times10^4$      & 6.13  & --    & --   \\
      \textbf{Current study}                           & \textbf{Exp.} & \textbf{Four merged jets} & \boldmath{$3.0\times10^4$}      & \textbf{6.02}  & \textbf{0.088} & \textbf{3.4}  \\
      \citet{BuchwaldRoundJet}                  & Exp. & Single free jet & $3.3\times10^4$      & 6.4~  & 0.103 & --   \\
      \citet{RodiRoundJet}                             & Exp. & Single free jet & $8.7\times10^4$      & 5.90  & 0.086 & --   \\
      \citet{WygnanskiPreservingJet}                 & Exp. & Single free jet & $\sim 10^5$          & 5.70  & 0.086 & 3.0  \\
      \citet{HusseinRoundJet}                  & Exp. & Single free jet & $\sim 10^5$          & 5.80  & 0.094 & 4.0  \\
      \citet{BauersfeldRoboticsMeetsFluidDynamics}                & Exp. & Four merged jets & --                   & 10.11* & 0.076 & 5.8  \\
      \hline
  \end{tabular}
  \caption{Comparison of round-jet scaling constants organized by $\Rey$ for the present diagonal-cut wake (bold) against canonical single-nozzle, twin-jet, and four-merged-jet configurations. ``Exp.'' and ``Sim.'' denote experimental and numerical simulation studies, respectively. *\,\citet{BauersfeldRoboticsMeetsFluidDynamics} report a fitted product $Bd = 10.11$\,mm for their unified multi-drone scaling, with $d$ taken as their jet exit diameter; this value is included here for reference but is not directly comparable due to the different normalization convention.}
  \label{tab:scaling_constants}
  \end{center}
\end{table}

The same scaling also establishes a Reynolds number for the merged flow. Evaluated using the  centerline velocity at the merge plane $ u_c(z_\text{merge}) \approx 4.24\,\si{m/s}$, and the effective source diameter $D_\text{eff} \approx 105\,\si{mm}$,
\begin{equation}
    \Rey_{D_\text{eff}} = \frac{ u_c(z_\text{merge})\,D_\text{eff}}{\nu} \approx 3.0\times 10^4,
    \label{eq:Re_Deff}
\end{equation}
which serves as a parameter for comparison with canonical studies, where the Reynolds number is conventionally associated with nozzle exit states.

\begin{figure}
  \centering
  \begin{subfigure}[b]{0.48\linewidth}
    \centering
    \includegraphics[width=\linewidth]{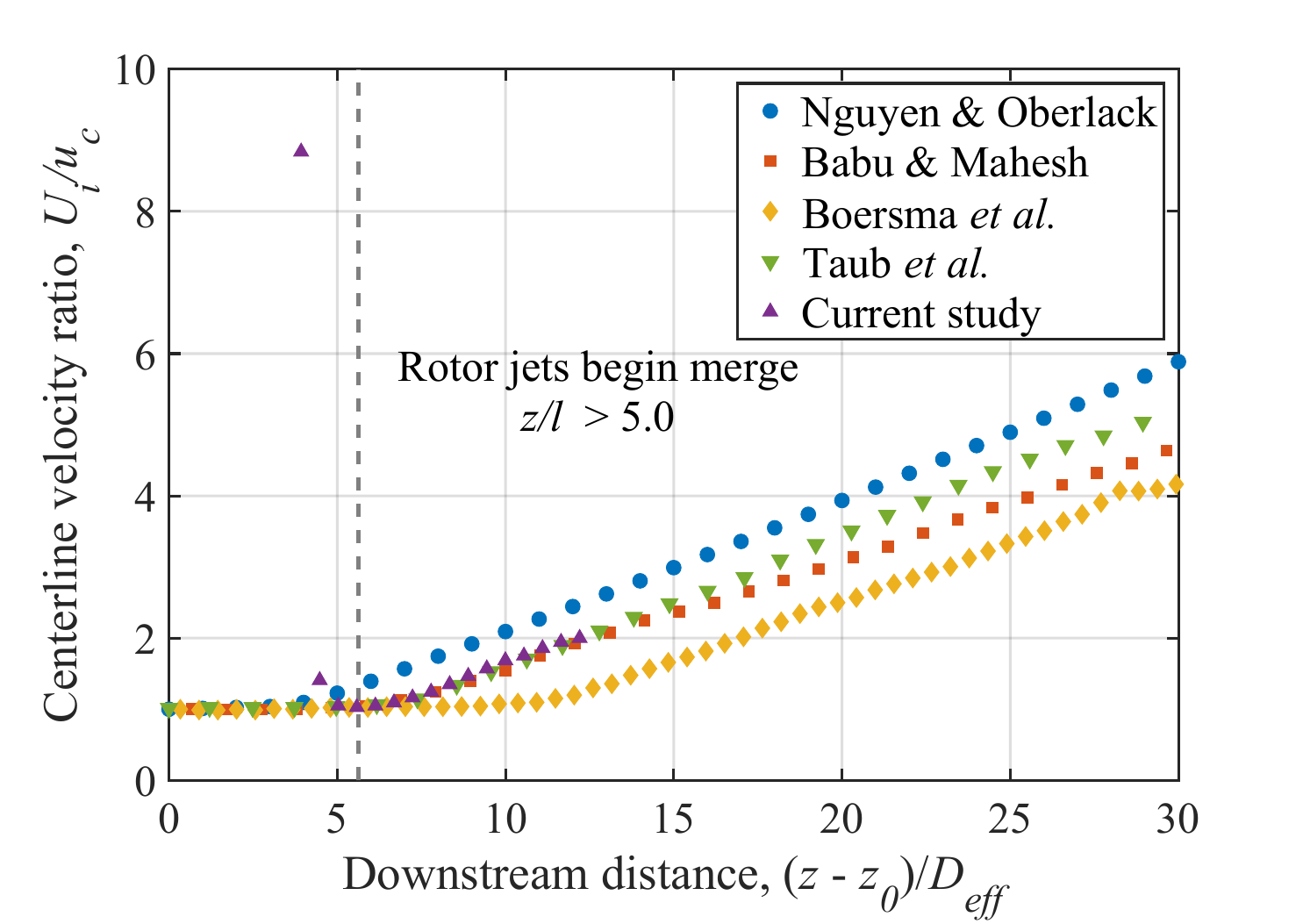}
    \caption{Centerline velocity decay comparison}
    \label{fig:decay_comparison}
  \end{subfigure}
  \hfill
  \begin{subfigure}[b]{0.48\linewidth}
    \centering
    \includegraphics[width=\linewidth]{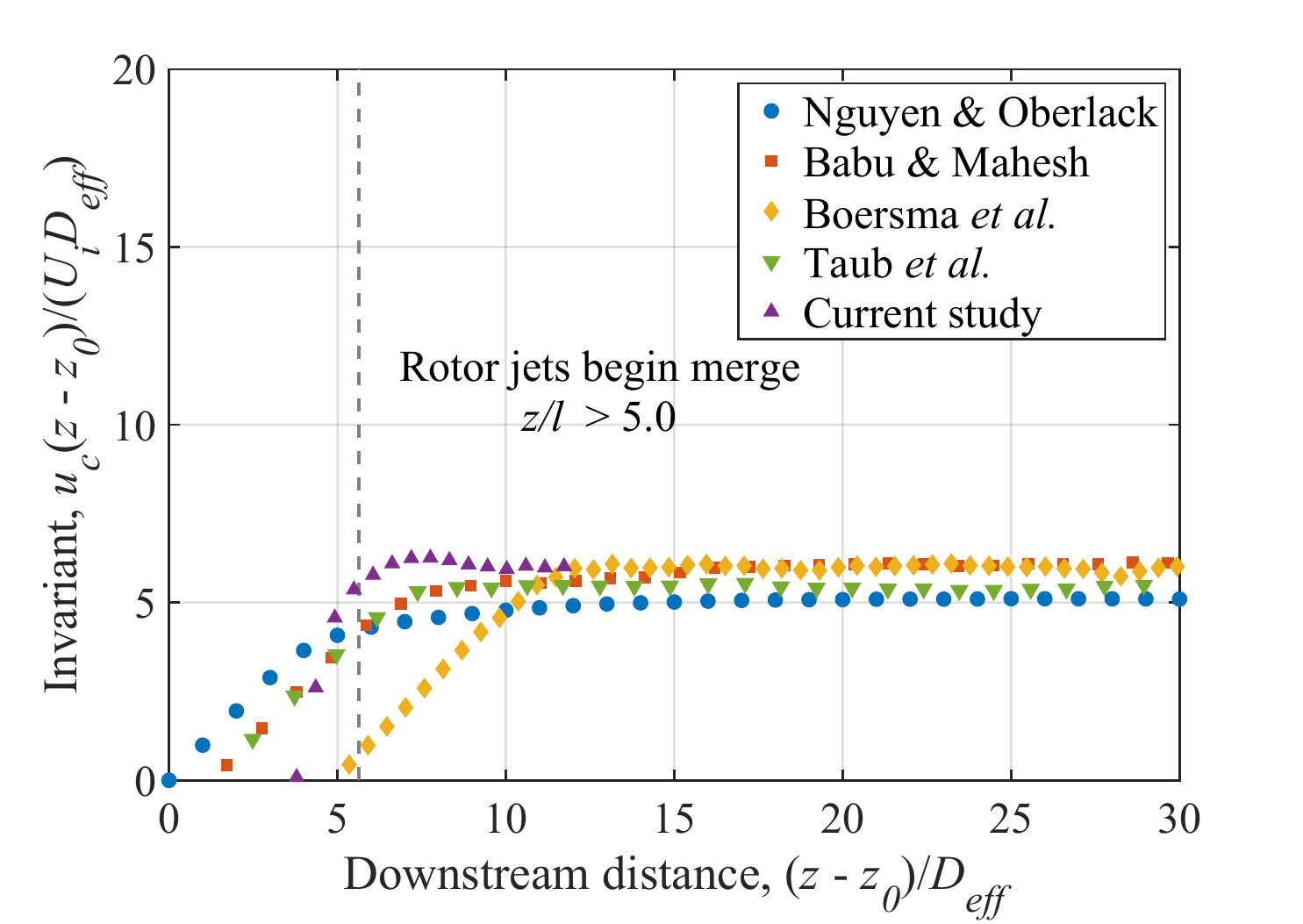}
    \caption{Self-similarity invariant}
    \label{fig:invariant_comparison}
  \end{subfigure}
  \caption{Far-field self-similarity of the merged diagonal-cut wake versus canonical round-jet studies. (\textit{a}) Inverse centerline velocity $U_i/u_c$ versus normalized downstream distance $(z - z_0)/D_\text{eff}$ showing linear decay beyond the merge plane ($B = 6.02$ and $|z_0|/D_\text{eff} = 3.42$). (\textit{b}) Self-similarity invariant $u_c(z - z_0)/(U_i D_\text{eff})$, plateauing near $6$. Vertical lines denote wake merging onset at $z/l = 5$. Reference data digitized from original sources:~\citet{NguyenDNSRoundJet, BabuDNSRoundJet, BoersmaJetDNS, TaubJetDNS}.}
  \label{fig:selfsimilar_farfield}
\end{figure}

Figure~\ref{fig:decay_comparison} compares the centerline decay against references in Table~\ref{tab:scaling_constants}, shifted by the virtual origin $|z_{0}|/D_{eff}$ to standardize the far-field origin convention across all datasets. Beyond the merge plane, the measurements align with reference data, confirming the merged wake is quantitatively consistent with the classical axisymmetric jet.

Rearranging equation~\eqref{eq:centerline_decay}, the invariant grouping
\begin{equation}
    \frac{u_c(z)\,(z - z_0)}{U_i\,D_\text{eff}} = B
    \label{eq:invariant}
\end{equation}
remains constant in self-similar flow regions. Beyond the merge plane, this invariant (Figure~\ref{fig:invariant_comparison}) plateaus near $6$, matching canonical round jets. Although the downstream extent is constrained by the field of view (\S\ref{sec:methodology}), the invariant profile remains flat, confirming that the merged wake achieves and sustains self-similarity.  

\subsection{Self-similarity of mean velocity}
\label{subsec:self_sim_mean_velocity}

\begin{figure}
  \centering
  \begin{subfigure}[b]{0.48\linewidth}
    \centering
    \includegraphics[width=\linewidth]{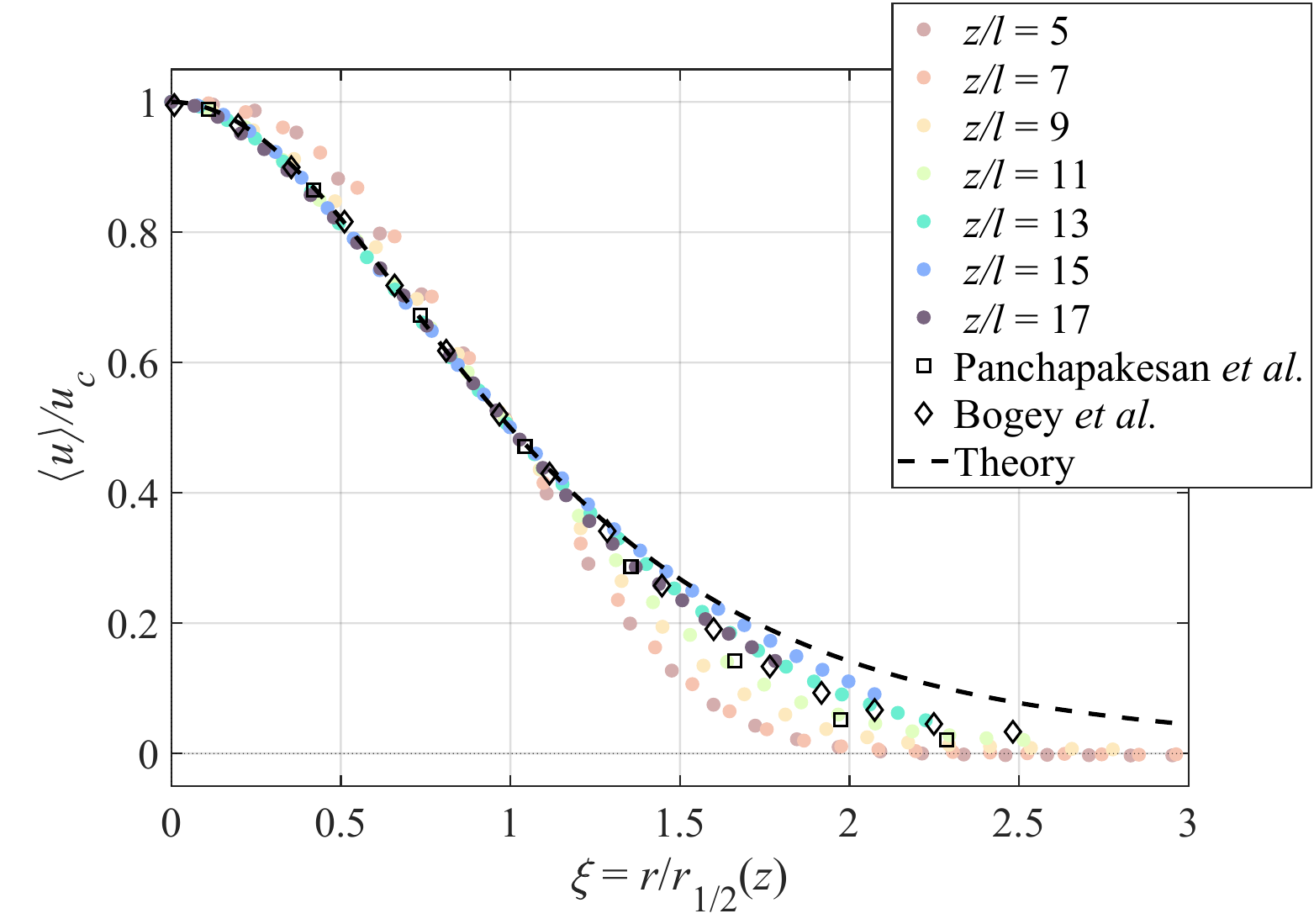}
    \caption{Streamwise velocity}
    \label{fig:self_sim_mean_u}
  \end{subfigure}
  \hfill 
  \begin{subfigure}[b]{0.48\linewidth}
    \centering
    \includegraphics[width=\linewidth]{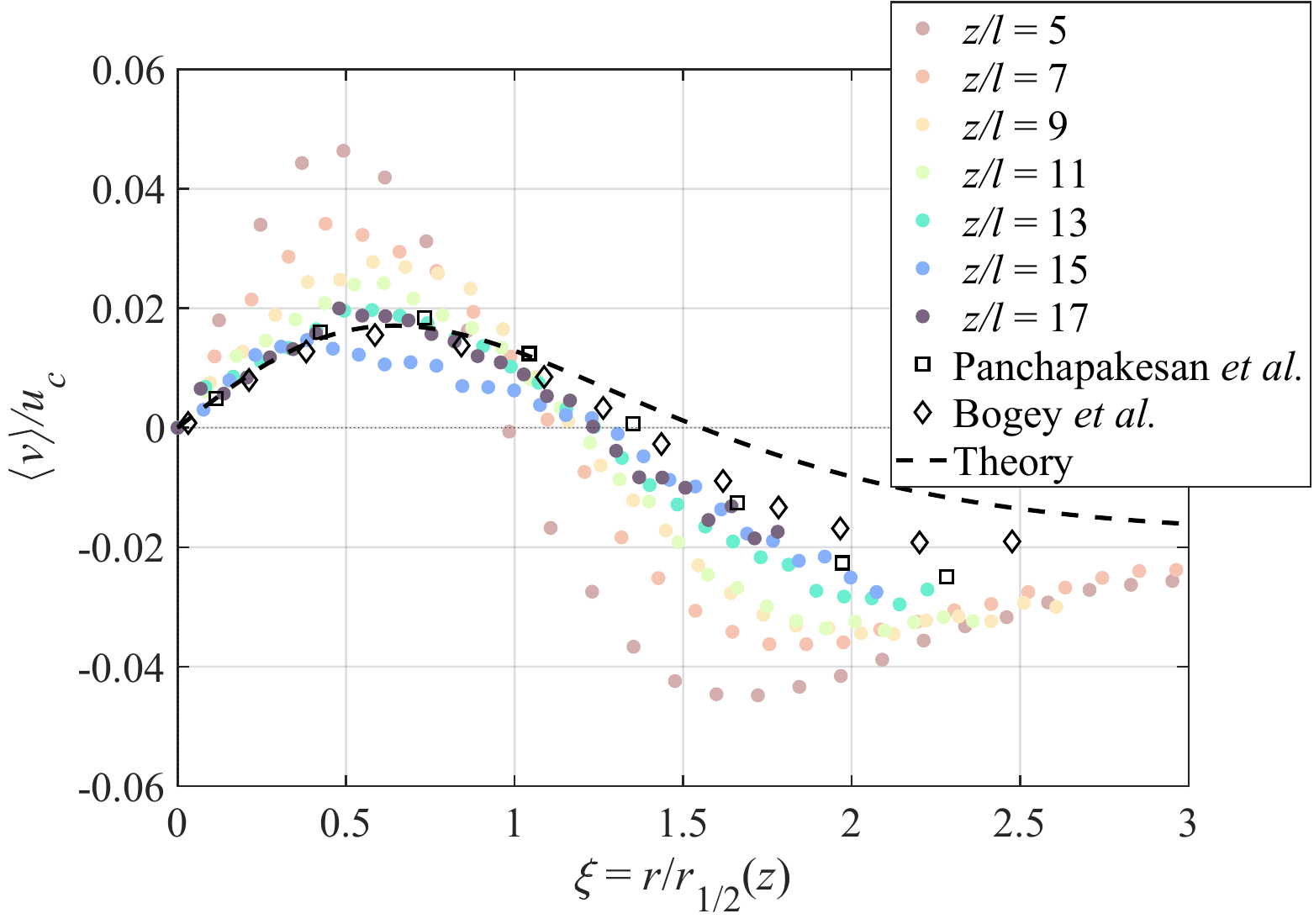}
    \caption{Cross-stream velocity}
    \label{fig:self_sim_mean_v}
  \end{subfigure}

  \caption{Self-similar mean velocity profiles in the far-field hovering quadrotor wake (diagonal cut). (\textit{a}) Streamwise $\langle u \rangle/u_c$ and (\textit{b}) cross-stream $\langle v \rangle/u_c$ velocity versus similarity variable $\xi = r/r_{1/2}(z)$ for $z/l \in \{5, 7, 9, 11, 13, 15, 17\}$. Profiles collapse for $z/l \gtrsim 13$, matching canonical round-jet behavior. Marker color encodes downstream position $z/l$, moving from warm to cool tones downstream. Digitized reference data are shown for experiment~(\citet{PanchapakesanAxisymmetricJet}, squares) and simulation~(\citet{BogeyRoundJetLES}, diamond). Dashed curves denote analytical self-similar solutions~\eqref{eq:axial_similarity} in (\textit{a}) and~\eqref{eq:lateral_similarity} in (\textit{b}), respectively.}
  \label{fig:self_sim_mean}
\end{figure}

The diagonal cut plane, where the near-field bimodal structure smooths by $z/l \approx 5$, is used to evaluate the merged jet's self-similarity. Profiles are normalized by local centerline velocity $u_c(z)$ and the similarity coordinate $\xi = r/r_{1/2}(z)$, where the half-width grows linearly, $r_{1/2}(z) = S\,(z - z_0)$, with spreading rate $S = 0.088$ and virtual origin $z_{0}$ from the half-width fit (\S\ref{subsec:centerline_scaling}, equation~\eqref{eq:half_width_def}). Assuming uniform turbulent transport, streamwise momentum balance yields the classical solution~\citep{PopeTurbulentFlows},
\begin{equation}
    \frac{\langle u\rangle(\xi)}{u_c(z)} \;=\; \frac{1}{\left[\,1 + \left(\sqrt{2}-1\right)\xi^{2}\,\right]^{2}}\,,
    \label{eq:axial_similarity}
\end{equation}
and continuity gives the cross-stream velocity,
\begin{equation}
    \frac{\langle v\rangle(\xi)}{u_c(z)} \;=\; \frac{S}{2}\,\frac{\xi - \left(\sqrt{2}-1\right)\xi^{3}}{\left[\,1 + \left(\sqrt{2}-1\right)\xi^{2}\,\right]^{2}}\,.
    \label{eq:lateral_similarity}
\end{equation} 

Streamwise profiles (figure~\ref{fig:self_sim_mean_u}) collapse onto equation~\eqref{eq:axial_similarity}, decaying to near zero by $\xi \approx 2.5$. The near-field stations ($z/l=5$--$11$) reflect a developing state before recovering classical equilibrium, while the far-field stations ($z/l= 13$--$17$) collapse tightly, matching reference data~\citet{PanchapakesanAxisymmetricJet, BogeyRoundJetLES}.

Cross-stream velocity profiles (figure~\ref{fig:self_sim_mean_v}) reproduce the canonical self-similar form of a free round jet, where the lateral spreading is associated with the inner positive peak, a zero crossing, and a negative outer region set by entrainment of ambient fluid. Far-field stations collapse around peak amplitude $\langle v\rangle/u_c \approx 0.018$--$0.020$, agreeing with equation~\eqref{eq:lateral_similarity} and reference studies.

For $\xi \gtrsim 1$, both components fall below analytical curves, tracking reference data. This departure stems from boundary intermittency between turbulent and ambient fluid, where reduced turbulent transport steepens the outer gradient relative to constant-transport assumptions~\citet{PopeTurbulentFlows}. 

Together, these first-order collapses establish that the mean downwash of a hovering quadrotor reaches mean-flow self-similarity in the far-field, validating canonical scaling laws as a baseline for evaluating higher-order turbulent moments.

\subsection{Fluctuating velocity fields}
While the mean flow achieves canonical self-similarity in the far-field, the turbulent Reynolds stresses develop more slowly and retain cut-dependent differences further downstream. The following profiles assess where, and how completely, the turbulent field approaches self-similarity. 

\subsubsection{Streamwise Reynolds normal stress}
\label{subsubsec:streamwise_Re_normal_stress}

\begin{figure}[htbp]
  \centering
  
  \begin{subfigure}[b]{0.48\linewidth}
    \centering
    \begin{overpic}[width=\linewidth]{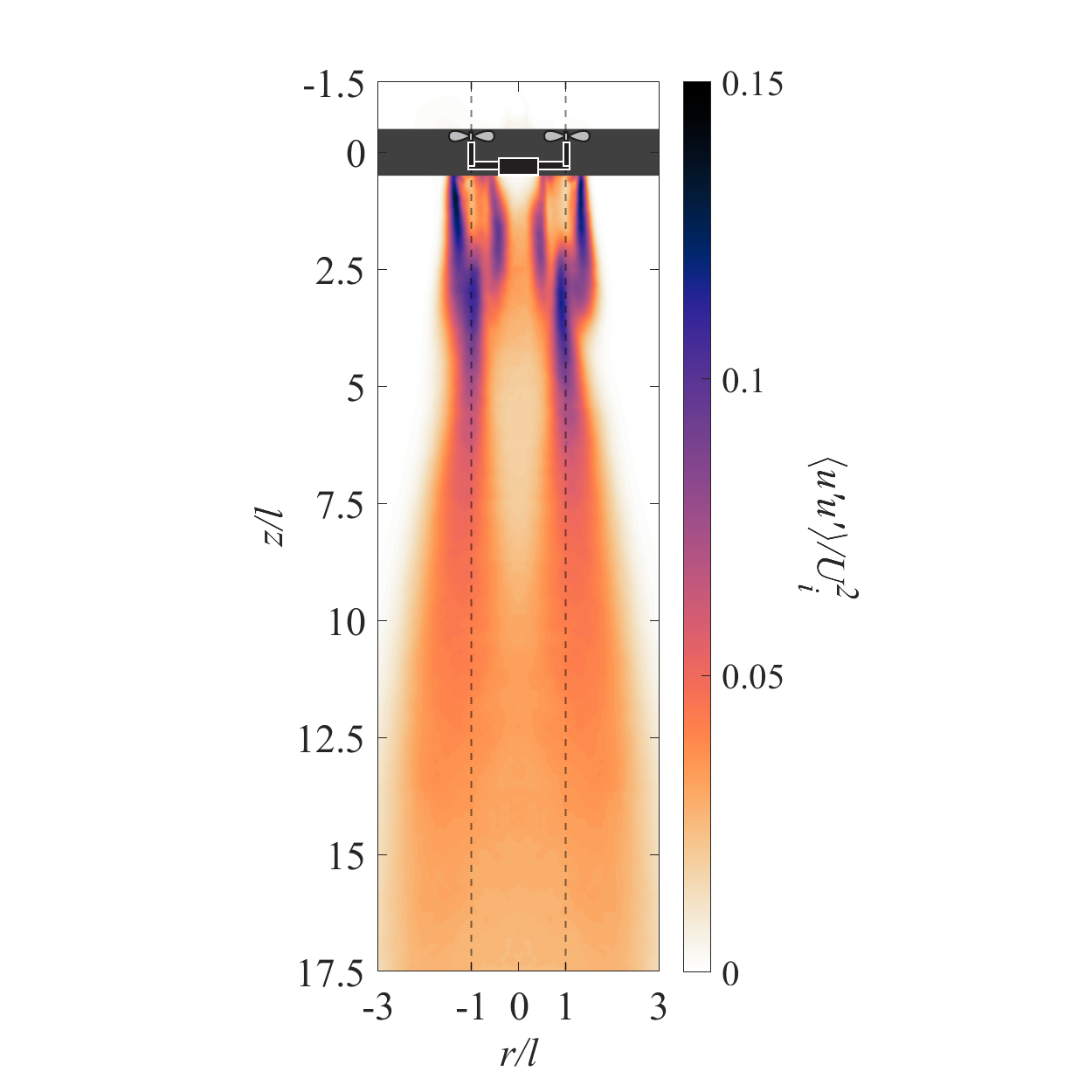}
    \put(75, 75){\includegraphics[width=0.25\linewidth]{figures/PDF_files/diagonal_cut_icon.pdf}}
    \end{overpic}
    \caption{Diagonal cut, $\langle u'u' \rangle$ contour}
    \label{fig:uu_fluc_contour_diag}
  \end{subfigure}
  \hfill
  \begin{subfigure}[b]{0.48\linewidth}
    \centering
    \begin{overpic}[width=\linewidth]{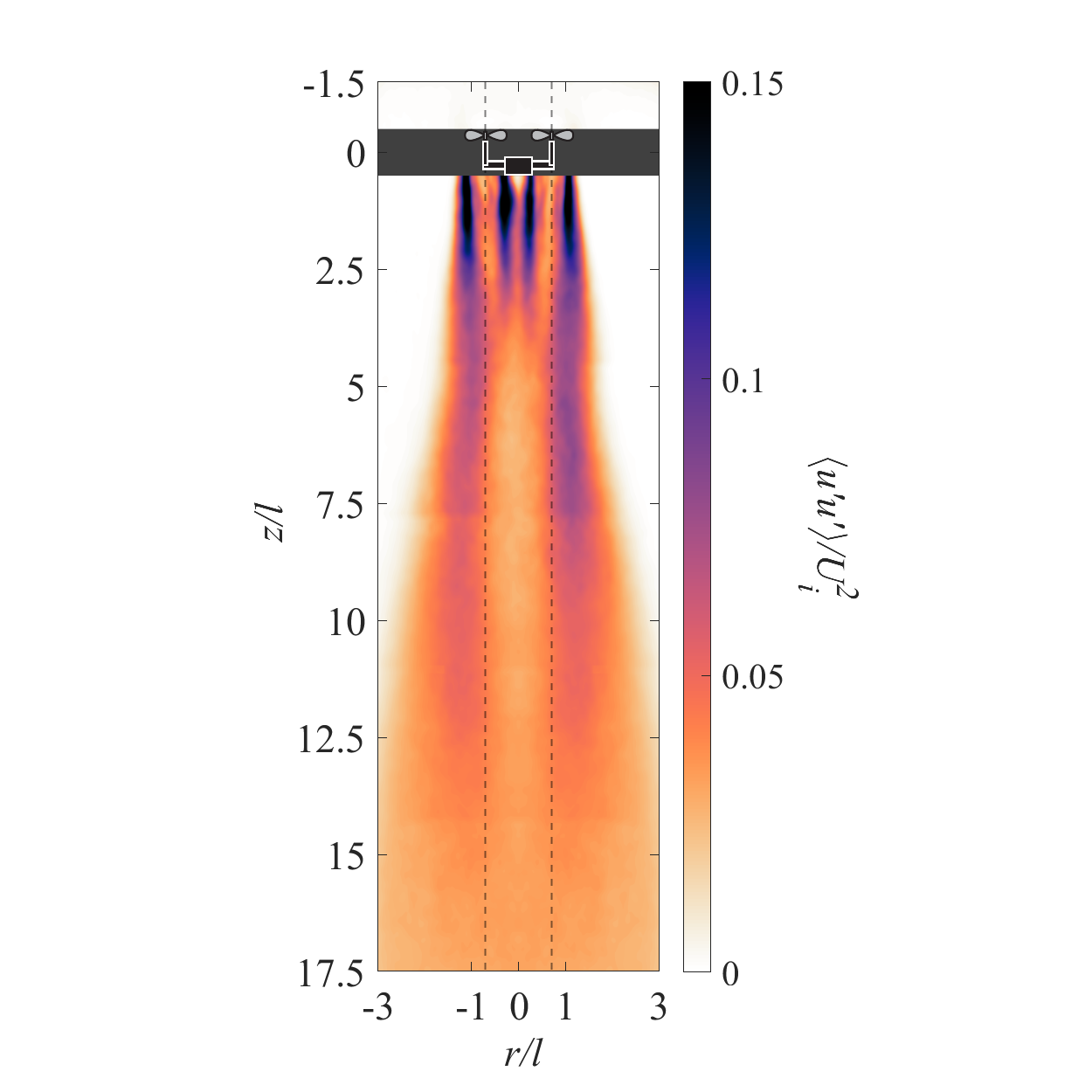}
    \put(75, 78.5){\includegraphics[width=0.25\linewidth]{figures/PDF_files/front_cut_icon.pdf}}
    \end{overpic}
    \caption{Front-rotor cut, $\langle u'u' \rangle$ contour}
    \label{fig:uu_fluc_contour_front}
  \end{subfigure}

  \vspace{6pt} 
  \begin{subfigure}[b]{0.48\linewidth}
    \centering
    \includegraphics[width=\linewidth]{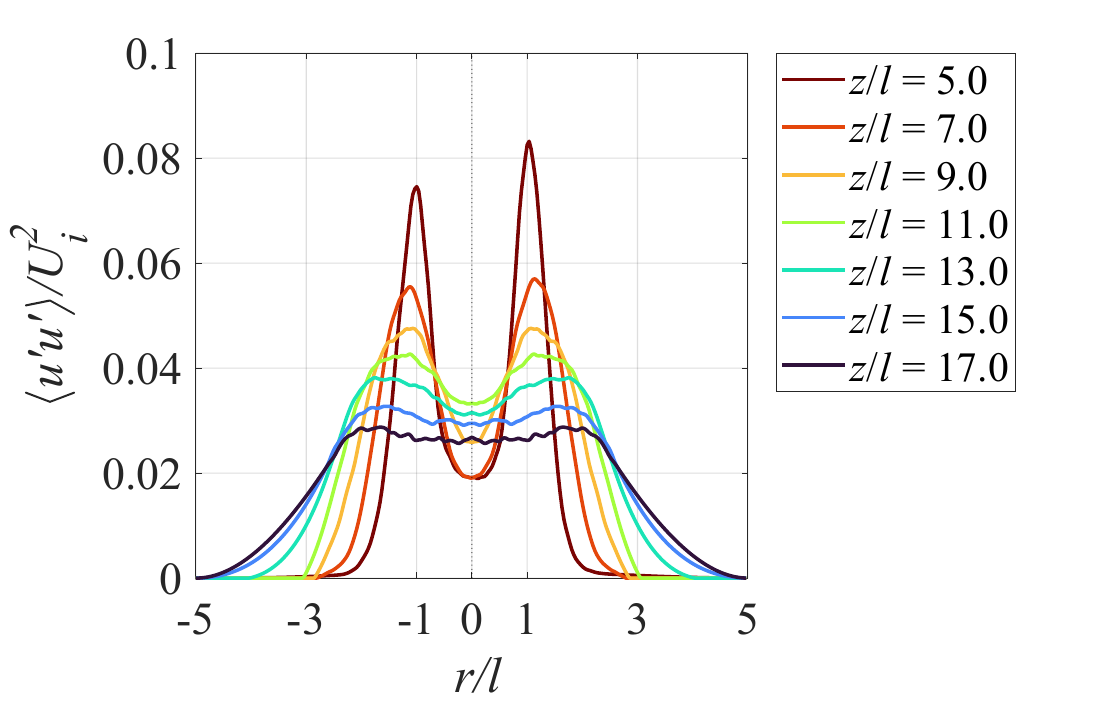}
    \caption{Diagonal cut, $\langle u'u' \rangle$ profiles}
    \label{fig:uu_fluc_profile_diag}
  \end{subfigure}
  \hfill
  \begin{subfigure}[b]{0.48\linewidth}
    \centering
    \includegraphics[width=\linewidth]{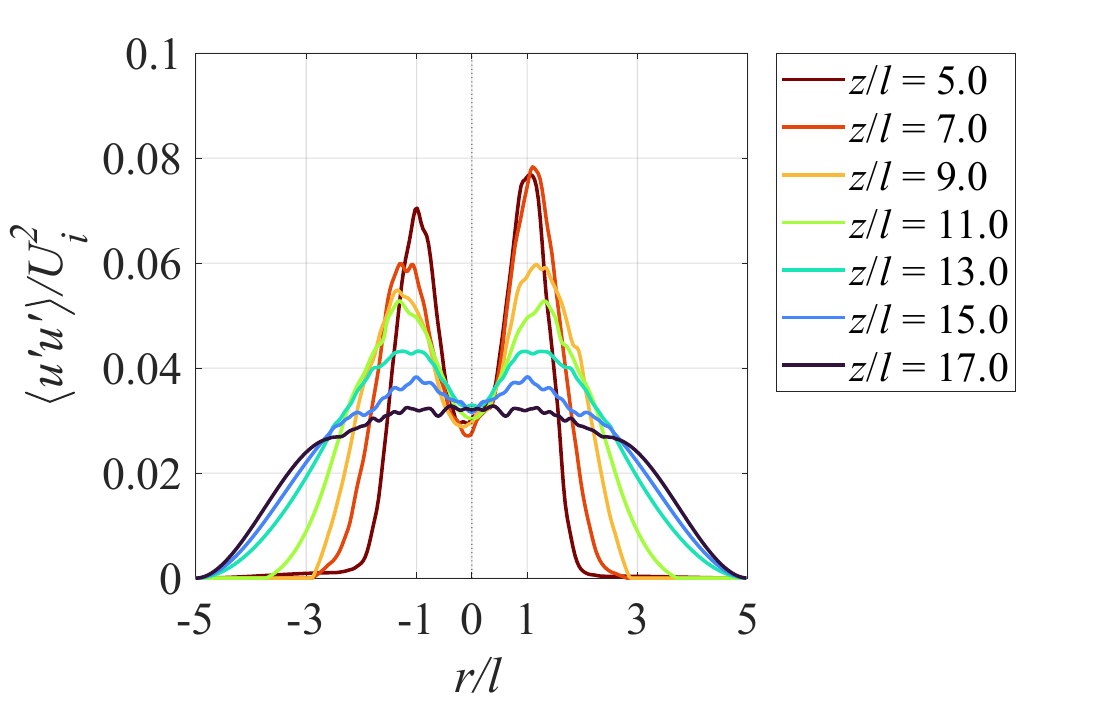}
    \caption{Front-rotor cut, $\langle u'u' \rangle$ profiles}
    \label{fig:uu_fluc_profile_front}
  \end{subfigure}

  \caption{Streamwise turbulent normal stress $\langle u'u' \rangle/U_i^2$ in the hovering quadrotor downwash. Contours: (\textit{a}) diagonal cut (plane A) and (\textit{b}) front-rotor cut (plane B). Dashed vertical lines mark intersected rotor positions, inset icons show plane orientation, and gray patches mask the quadrotor body. Profiles: (\textit{c}, \textit{d}) cross-stream distributions at downstream stations $z/l \in \{5, 7, 9, 11, 13, 15, 17\}$, where cooler colors indicate locations farther downstream.}
  \label{fig:uu_Re_normal_stresses}
\end{figure}

Although the mean velocity field merges into a single quasi-axisymmetric jet, the streamwise normal stress $\langle u'u'\rangle/U_i^2$ retains a persistent multi-source signature (figures~\ref{fig:uu_fluc_contour_diag} and~\ref{fig:uu_fluc_contour_front}). In the near-field, stress concentrates in narrow stripes along the high-gradient shear-layer boundaries of each rotor jet. In the diagonal cut, wider rotor separation prevents inner shear layers from merging immediately, preserving distinct off-axis stress bands.

Radial profiles (figures~\ref{fig:uu_fluc_profile_diag} and~\ref{fig:uu_fluc_profile_front}) show this bimodal structure persisting downstream. At $z/l = 5$, both cuts exhibit twin off-axis peaks and a centerline deficit that is deeper in the diagonal cut. As the wake develops, the profiles broaden and flatten into a central plateau. Closer spacing in the front-cut promotes earlier inner shear-layer interaction along the centerline, analogous to canonical twin jets~\citep{LabanTwinJetsExp, HarimaTwinJets}. Conversely, wider separation in the diagonal cut delays this interaction, retaining turbulence memory far downstream. 

This persistence of bimodal stress structure beyond the merging of the mean velocity is the characteristic signature of multi-source jet flows. This has been documented in dual planar jets~\citep{TanakaParallelJets, MillerDualJet, LinParallelJets}, twin circular jets \citep{OkamotoTurbulentJet, HarimaTwinJets}, and resolved more recently with PIV and LES \citep{NasrParallelJets, AndersonParallelJets, LeeParallelJets, LiTwinJets}. Across these flows, the mean-flow self-similarity establishes significantly earlier than second-moment equilibrium. Our data confirms that this scaling holds for the four-rotor quadrotor wake, where the streamwise normal stress preserves a bimodal multi-rotor signature even after the mean velocity profile has begun to merge into a single column.

\begin{figure}
  \centering
  
  \begin{subfigure}[b]{0.48\linewidth}
    \centering
    \begin{overpic}[width=\linewidth]{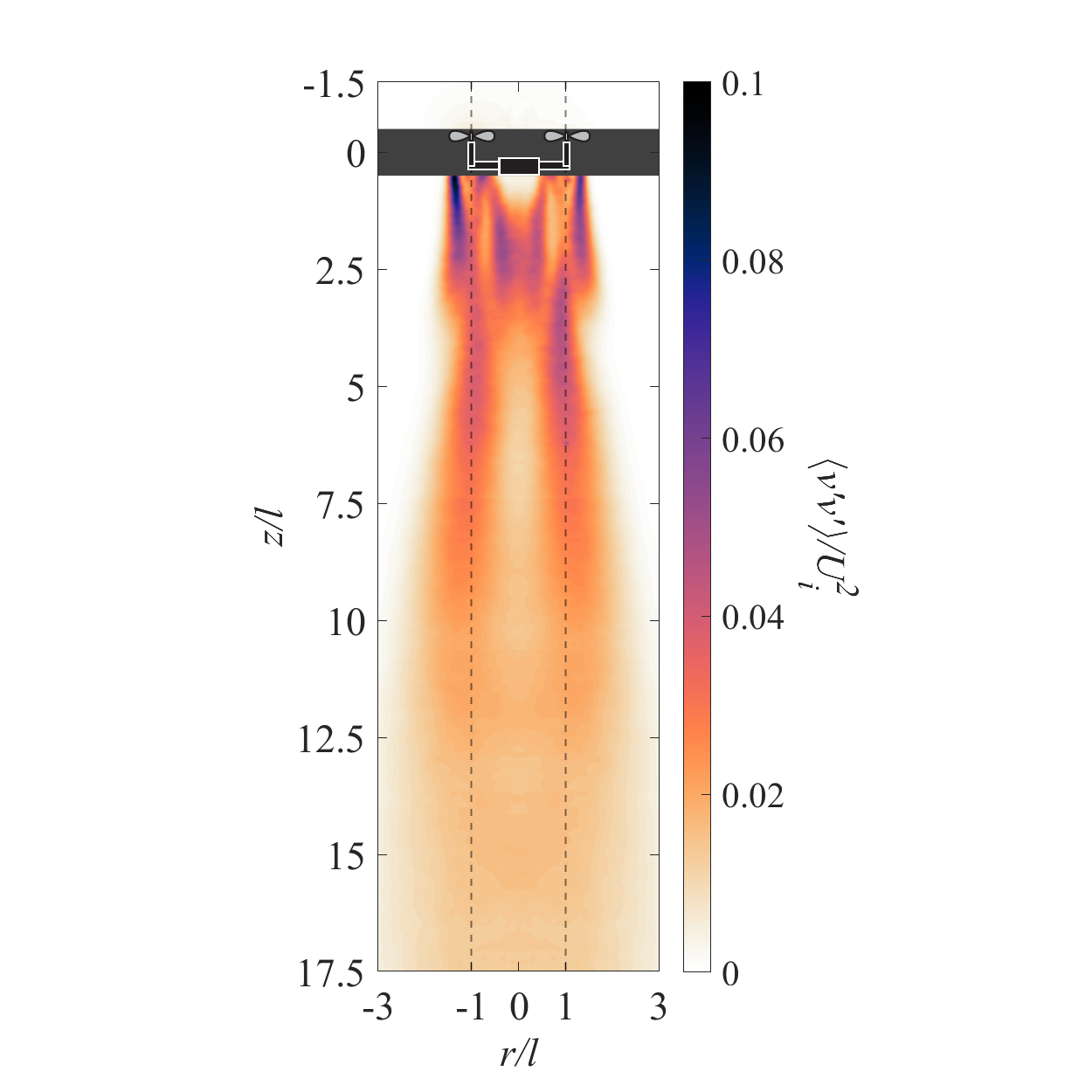}
    \put(75, 75){\includegraphics[width=0.25\linewidth]{figures/PDF_files/diagonal_cut_icon.pdf}}
    \end{overpic}
    \caption{Diagonal cut, $\langle v'v' \rangle$ contour}
    \label{fig:vv_fluc_contour_diag}
  \end{subfigure}
  \hfill
  \begin{subfigure}[b]{0.48\linewidth}
    \centering
    \begin{overpic}[width=\linewidth]{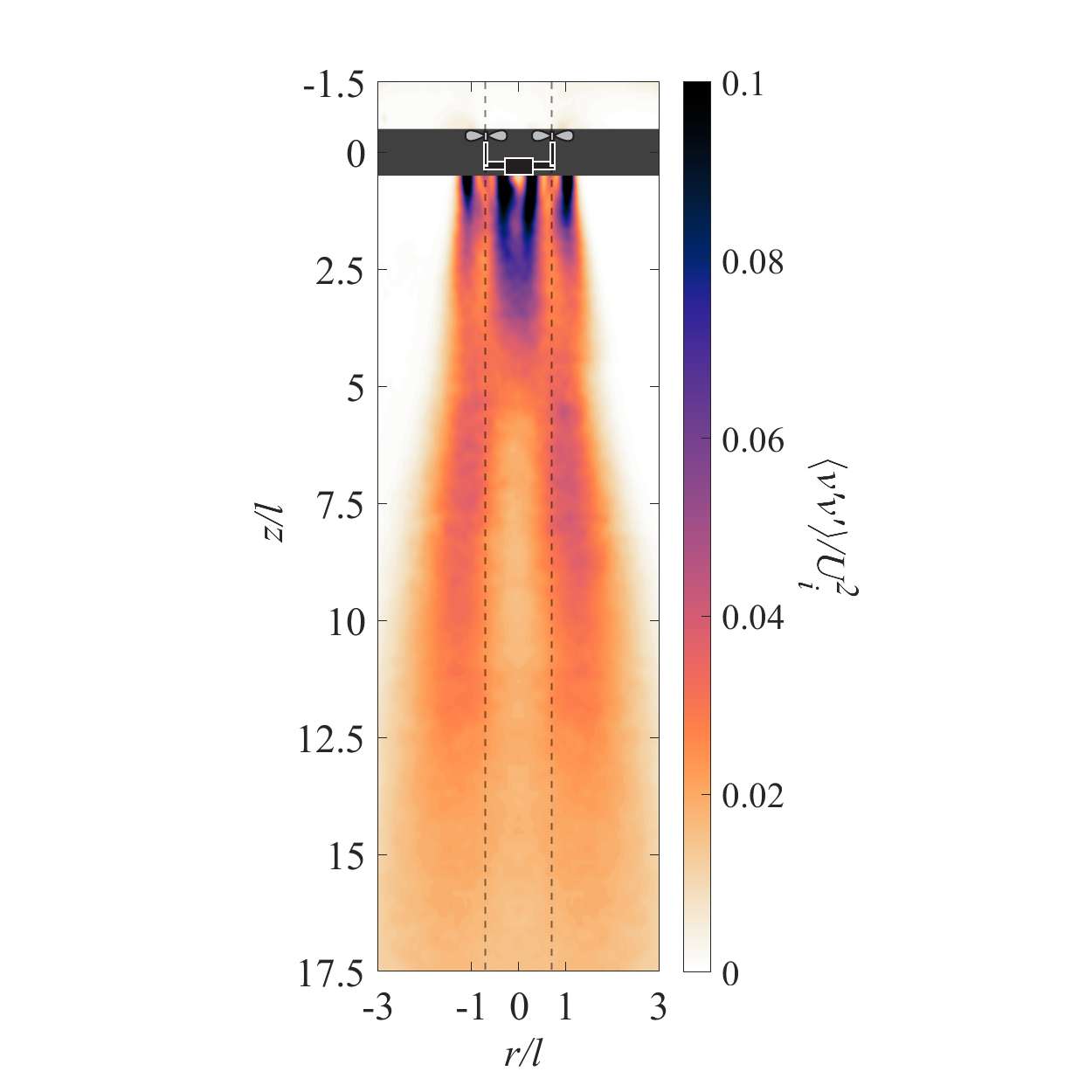}
    \put(75, 78.5){\includegraphics[width=0.25\linewidth]{figures/PDF_files/front_cut_icon.pdf}}
    \end{overpic}
    \caption{Front-rotor cut, $\langle v'v' \rangle$ contour}
    \label{fig:vv_fluc_contour_front}
  \end{subfigure}

  \vspace{6pt} 
  \begin{subfigure}[b]{0.48\linewidth}
    \centering
    \includegraphics[width=\linewidth]{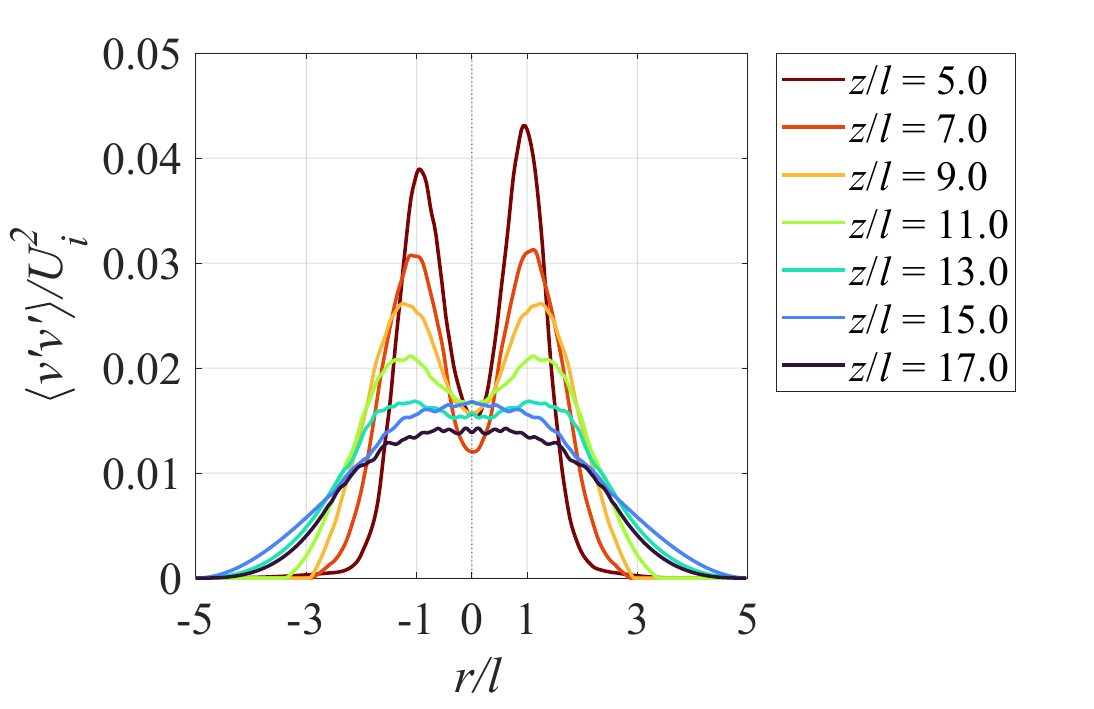}
    \caption{Diagonal cut, $\langle v'v' \rangle$ profiles}
    \label{fig:vv_fluc_profile_diag}
  \end{subfigure}
  \hfill
  \begin{subfigure}[b]{0.48\linewidth}
    \centering
    \includegraphics[width=\linewidth]{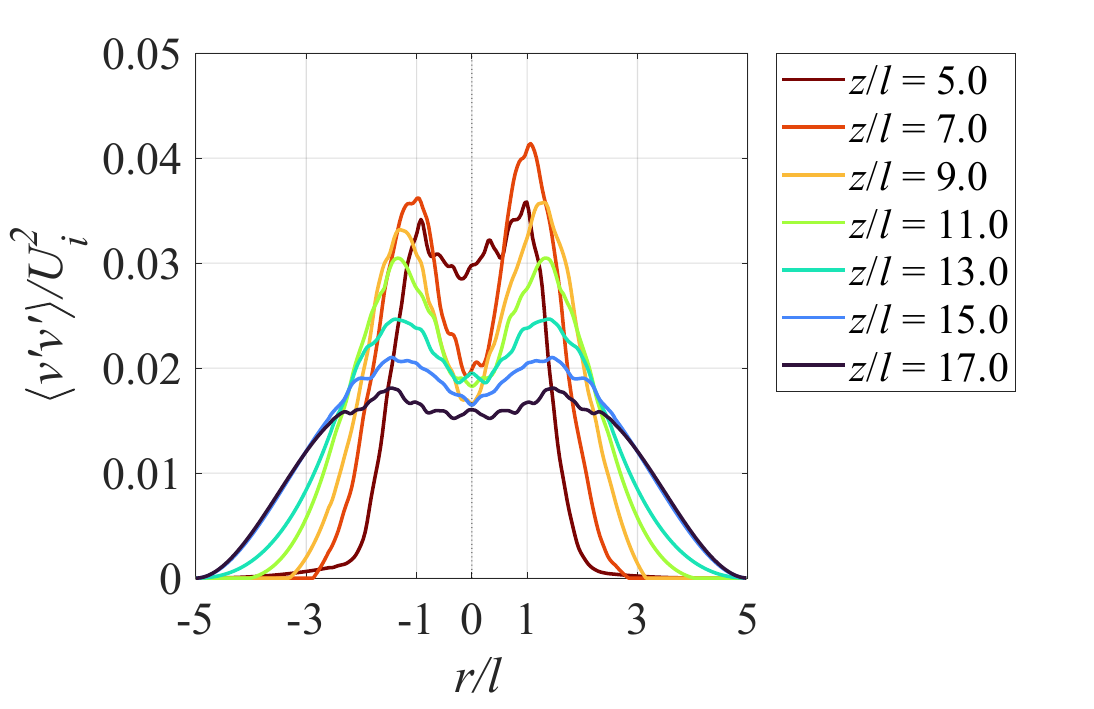}
    \caption{Front-rotor cut, $\langle v'v' \rangle$ profiles}
    \label{fig:vv_fluc_profile_front}
  \end{subfigure}

  \caption{Cross stream turbulent normal stress $\langle v'v' \rangle/U_i^2$ in the hovering quadrotor downwash. Contours: (\textit{a}) diagonal cut (plane A) and (\textit{b}) front-rotor cut (plane B). Dashed vertical lines mark intersected rotor positions, inset icons show plane orientation, and gray patches mask the quadrotor body. Profiles: (\textit{c}, \textit{d}) cross-stream distributions at downstream stations $z/l \in \{5, 7, 9, 11, 13, 15, 17\}$, where cooler colors indicate locations farther downstream.}
  \label{fig:vv_Re_normal_stresses}
\end{figure}

\subsubsection{Cross-stream Reynolds normal stress}
\label{subsubsec:cross-stream_Re_normal_stress}

The magnitude of the cross-stream Reynolds normal stress, $\langle v'v'\rangle/U_i^2$, reaches roughly half the magnitude of its streamwise counterpart and peaks off-axis, consistent with canonical round-jet literature; the cross-stream fluctuations peak off-axis, contrasting with the streamwise fluctuations, which reach their maximum along the centerline. The filled contours(figures~\ref{fig:vv_fluc_contour_diag} and~\ref{fig:vv_fluc_contour_front}) display persistent bands of elevated stress extending far downstream, particularly in the front-rotor cut where the bimodal structure remains visible to $z/l \approx 15$.

The two cuts evolve at markedly different rates (cf. figures~\ref{fig:vv_fluc_contour_diag} and~\ref{fig:vv_fluc_contour_front}). While canonical round jets exhibit delayed cross-stream self-similarity relative to streamwise stresses~\citep{HusseinRoundJet, PanchapakesanAxisymmetricJet}, the multi-rotor geometry exaggerates this lag. The diagonal cut centerline depression gradually closes toward a single peak, whereas the front-rotor cut preserves twin off-axis peaks throughout the domain up to $z/l = 17$  (cf. figures~\ref{fig:vv_fluc_profile_diag} and~\ref{fig:vv_fluc_profile_front}).

This prolonged structural memory in the front cut is driven by intense mean lateral velocities ($|\langle v \rangle|/U_i$); \S\ref{subsubsec:v_mean}) and strong inner shear layers where opposing rotor flows interact, generating localized turbulent kinetic energy along the rotor interface~\citep{ZhouRotorRotor}. Consequently, the cross-stream field retains the multi-rotor signature longer than the streamwise component (figure~\ref{fig:vv_Re_normal_stresses}).

Ultimately, while the normal stresses reveal the bimodal turbulence structure inherited from this multi-source origin, the Reynolds shear stress, in contrast, reveals where momentum is being rapidly transported across the wake.

\subsection{Reynolds shear stress}
\label{subsec:Re_shear_stress}
\begin{figure}[htbp]
  \centering
  
  \begin{subfigure}[b]{0.48\linewidth}
    \centering
    \begin{overpic}[width=\linewidth]{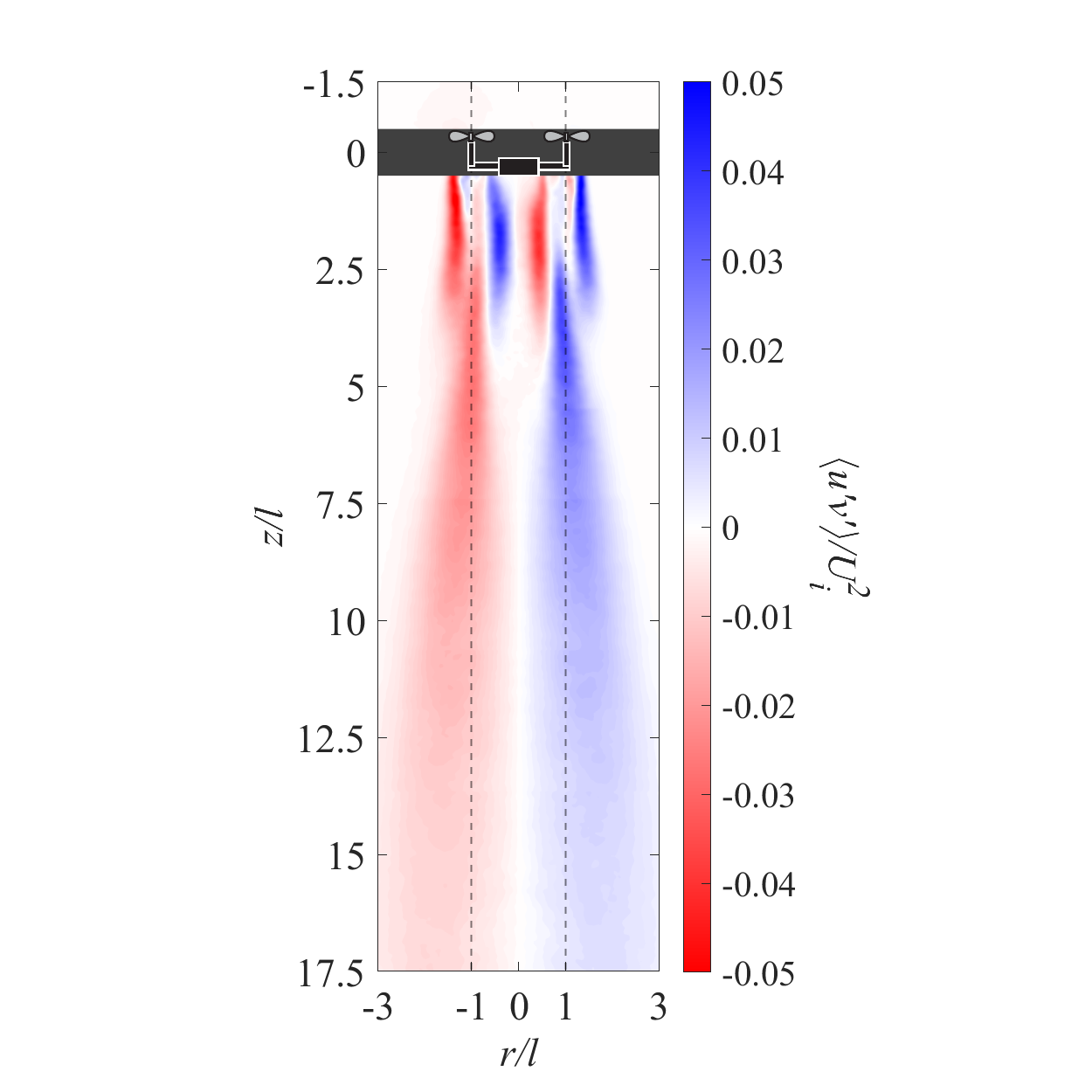}
    \put(75, 75){\includegraphics[width=0.25\linewidth]{figures/PDF_files/diagonal_cut_icon.pdf}}
    \end{overpic}
    \caption{Diagonal cut, $\langle u'v' \rangle$ contour}
    \label{fig:uv_fluc_contour_diag}
  \end{subfigure}
  \hfill
  \begin{subfigure}[b]{0.48\linewidth}
    \centering
    \begin{overpic}[width=\linewidth]{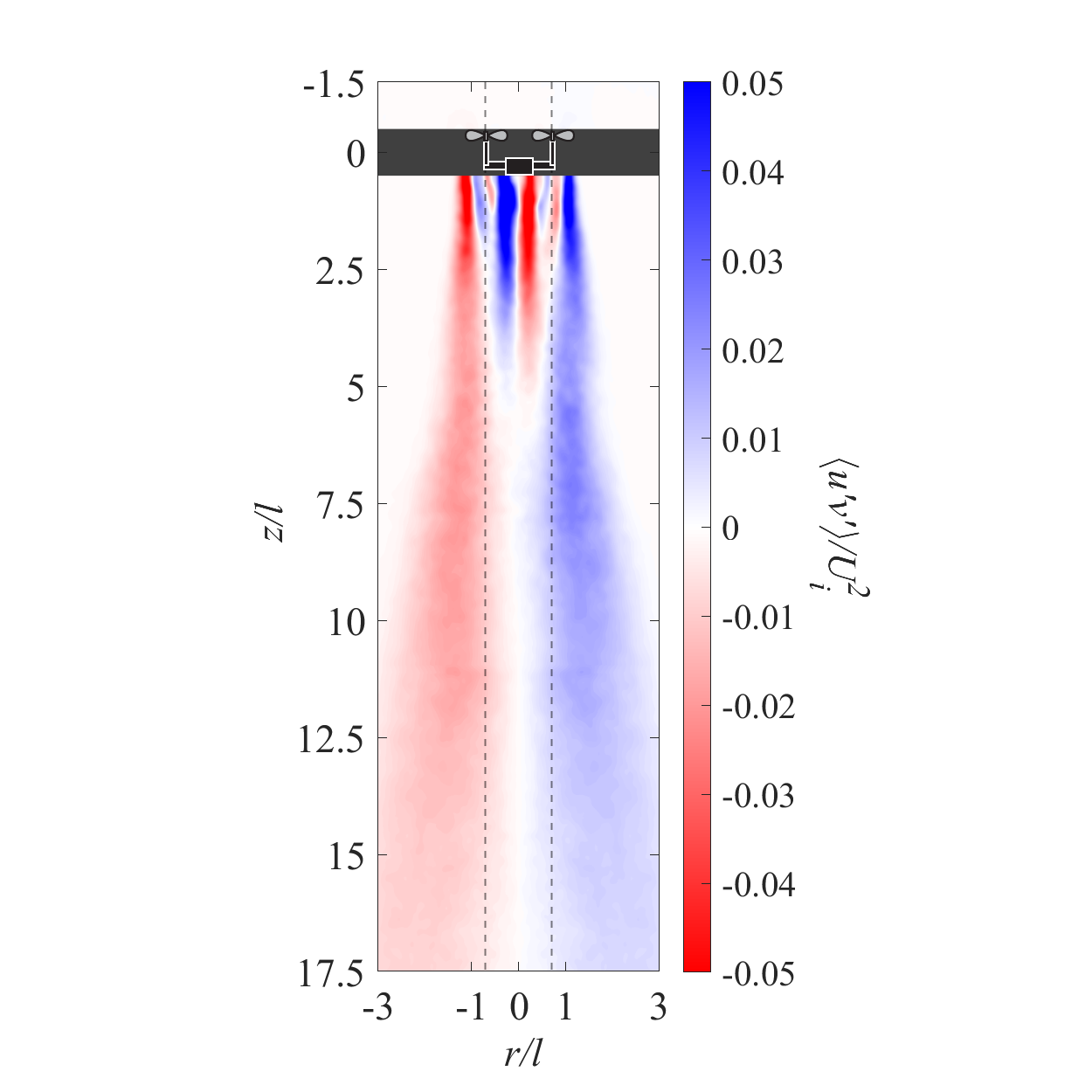}
    \put(75, 78.5){\includegraphics[width=0.25\linewidth]{figures/PDF_files/front_cut_icon.pdf}}
    \end{overpic}
    \caption{Front-rotor cut, $\langle u'v' \rangle$ contour}
    \label{fig:uv_fluc_contour_front}
  \end{subfigure}
  \vspace{6pt}  
  \begin{subfigure}[b]{0.48\linewidth}
    \centering
    \includegraphics[width=\linewidth]{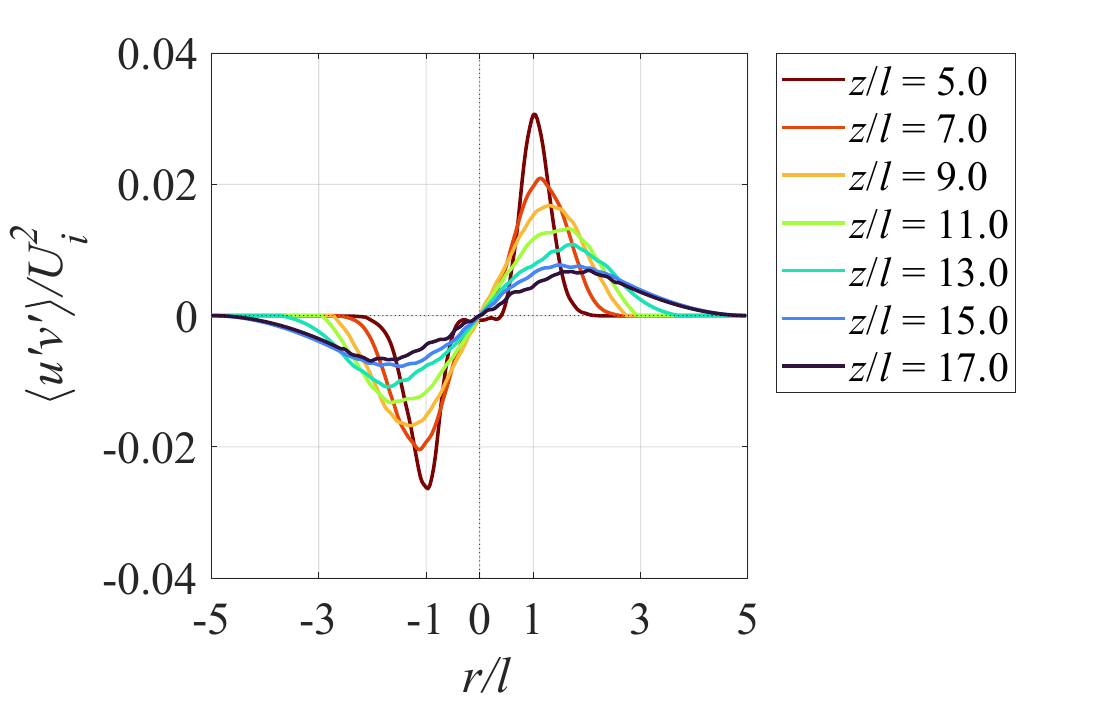}
    \caption{Diagonal cut, $\langle u'v' \rangle$ profiles}
    \label{fig:uv_fluc_profile_diag}
  \end{subfigure}
  \hfill
  \begin{subfigure}[b]{0.48\linewidth}
    \centering
    \includegraphics[width=\linewidth]{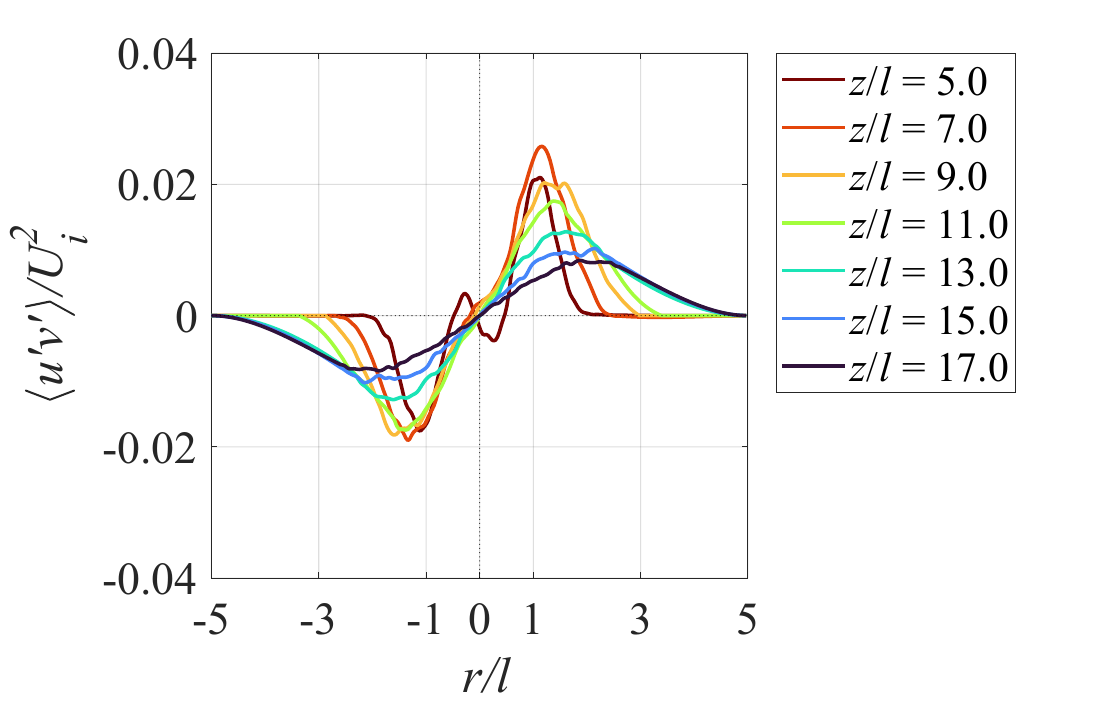}
    \caption{Front-rotor cut, $\langle u'v' \rangle$ profiles}
    \label{fig:uv_fluc_profile_front}
  \end{subfigure}

  \caption{Reynolds shear stress $\langle u'v' \rangle/U_i^2$ in hovering quadrotor downwash. Contours: (\textit{a}) diagonal cut (plane A) and (\textit{b}) front-rotor cut (plane B). Dashed vertical lines mark intersected rotor positions, inset icons show plane orientation, and gray patches mask the quadrotor body. Profiles: (\textit{c}, \textit{d}) cross-stream distributions at downstream stations $z/l \in \{5, 7, 9, 11, 13, 15, 17\}$, where cooler colors indicate locations farther downstream.}
  \label{fig:uv_Re_shear_stresses}
\end{figure}

The Reynolds shear stress, $\langle u'v'\rangle/U_i^2$, exhibits stronger resemblance to a canonical turbulent jet than either normal stress component (\S\ref{subsubsec:streamwise_Re_normal_stress} and \S\ref{subsubsec:cross-stream_Re_normal_stress}), mediating radial momentum transport and ambient entrainment. As shown in figures~\ref{fig:uv_fluc_contour_diag} and~\ref{fig:uv_fluc_contour_front}, both cuts display the classical antisymmetric structure ($\partial\langle u\rangle/\partial r \lessgtr 0$ for $r \gtrless 0$), predicted directly by eddy-viscosity closure,
\begin{equation}
    -\langle u'v'\rangle = \nu_T\,\frac{\partial \langle u\rangle}{\partial r},
    \label{eq:Re_shear_closure}
\end{equation}
where $\nu_T > 0$ is the local eddy viscosity~\citep{PopeTurbulentFlows}. 

In the diagonal cut (figure~\ref{fig:uv_fluc_profile_diag}), profiles remain antisymmetric across the domain, with peak magnitudes coinciding with the maximum radial gradient of $\langle u\rangle$ and weakening smoothly downstream. The front-rotor cut (cf. figures \ref{fig:uv_fluc_profile_diag} and \ref{fig:uv_fluc_profile_front}) displays near-field secondary peaks ($r/l \approx \pm 0.5$) from inner shear layers between closely spaced rotors. However, these multi-source features fade rapidly by $z/l \approx 7$, after which both cuts settle into the same canonical shear-stress distribution.

The faster collapse of $\langle u'v'\rangle$ relative to the normal stresses stems from its direct coupling to the mean velocity gradient in equation~\eqref{eq:Re_shear_closure}, and the turbulence production term $P = -\langle u'v'\rangle\,\partial\langle u\rangle/\partial r$. Once the mean flow merges into a smooth, single-peaked column (\S\ref{subsubsec:u_mean}), the shear stress rapidly adjusts to support it, tracking mean-flow evolution far more closely than $\langle u'u'\rangle$ or $\langle v'v'\rangle$.

\subsection{Self-similarity of turbulent stresses}
\label{subsec:self_sim_stresses}

\begin{figure}[htbp]
  \centering  
  \begin{subfigure}[b]{0.48\linewidth}
    \centering
    \includegraphics[width=\linewidth]{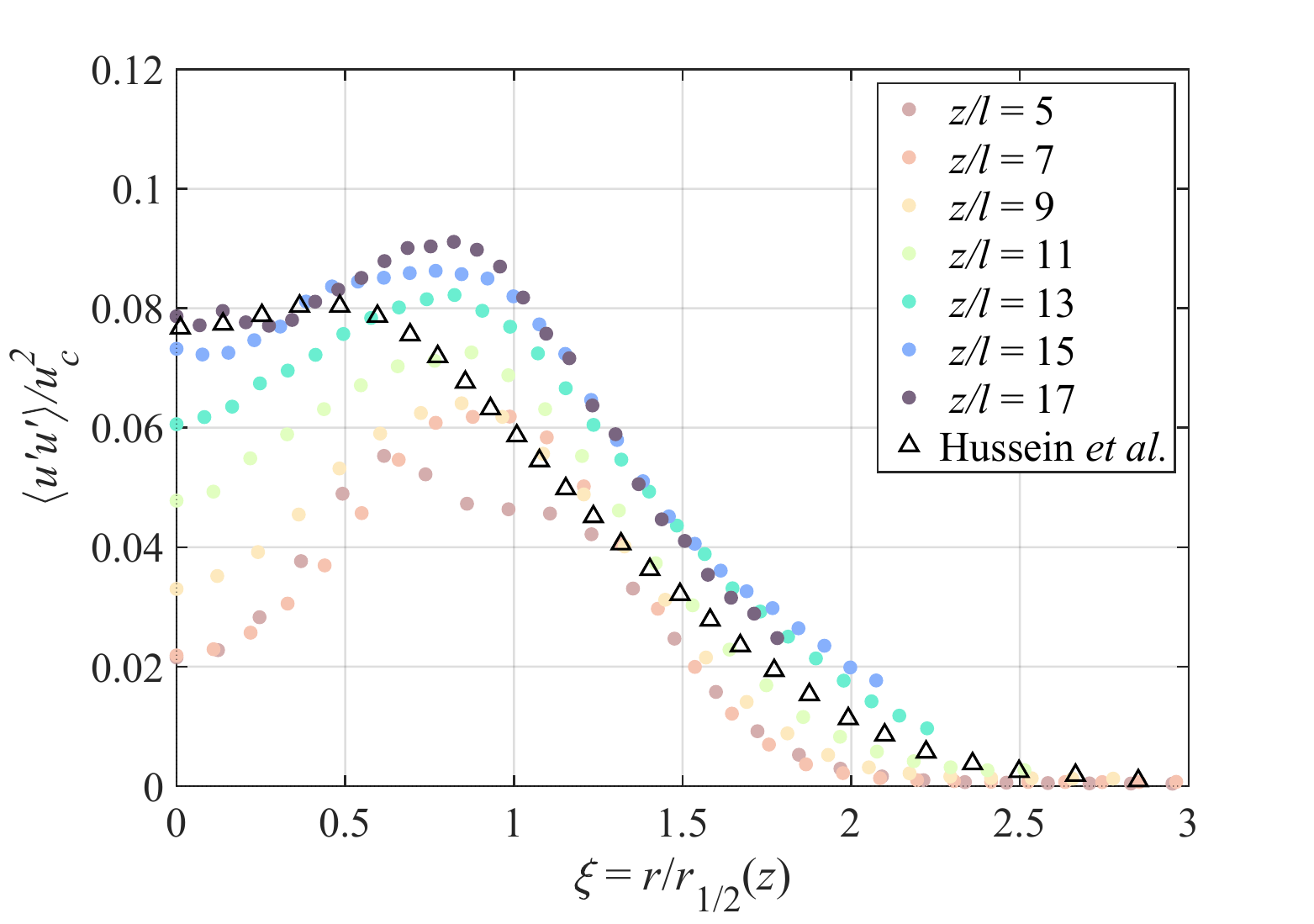}
    \caption{Streamwise turbulent normal stress}
    \label{fig:self_sim_fluc_uu}
  \end{subfigure}
  \hfill 
  \begin{subfigure}[b]{0.48\linewidth}
    \centering
    \includegraphics[width=\linewidth]{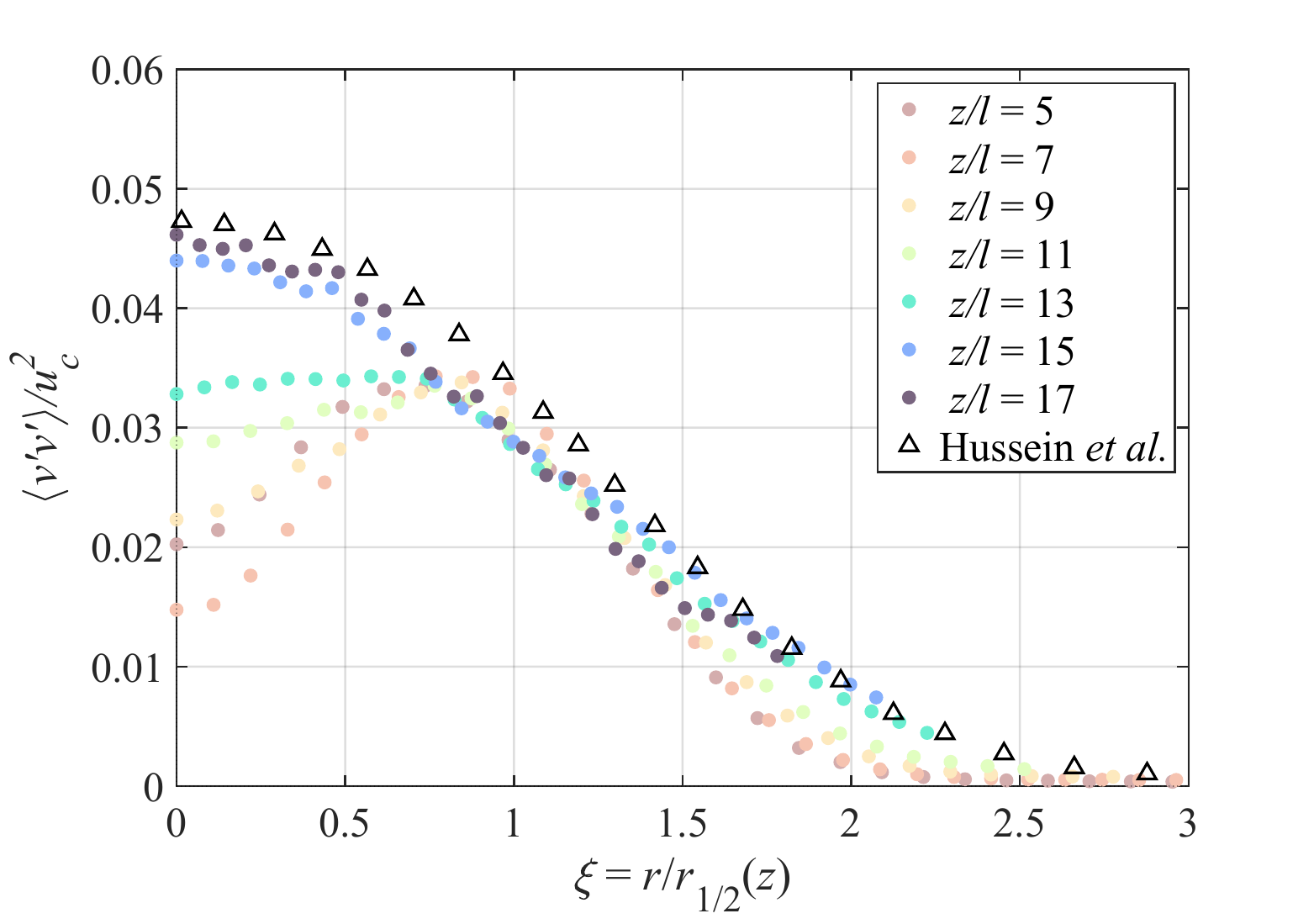}
    \caption{Cross-stream turbulent normal stress}
    \label{fig:self_sim_fluc_vv}
  \end{subfigure}
  \hfill
  \begin{subfigure}[b]{0.48\linewidth}
    \centering
    \includegraphics[width=\linewidth]{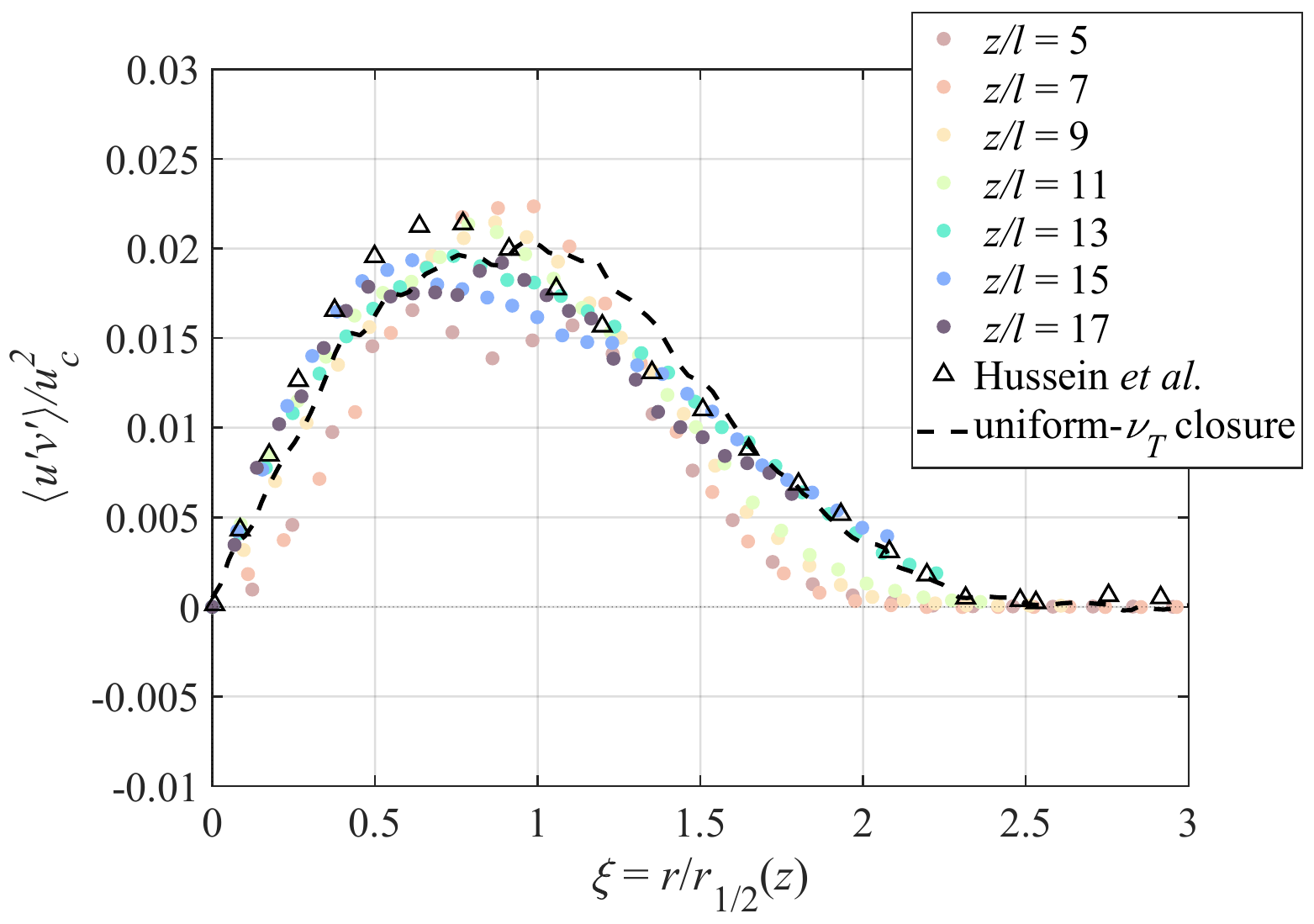}
    \caption{Reynolds shear stress}
    \label{fig:self_sim_fluc_uv}
  \end{subfigure}
  \caption{Self-similar Reynolds stresses for the diagonal cut, normalized by the square of centerline streamwise velocity $u_c^2$. Reference data digitized from~\citet{HusseinRoundJet} (upward-pointing triangle symbols), reported in variance form $\langle u_i'u_j'\rangle/u_c^2$, are shown for comparison. The black dashed curve in (\textit{c}) denotes uniform-eddy-viscosity closure with $\hat{\nu}_T = 0.028$.}
  \label{fig:self_sim_stresses}
\end{figure}

The streamwise turbulent normal stress $\langle u'u'\rangle/u_c^2$ gradually approaches a self-similar state (figure~\ref{fig:self_sim_fluc_uu}). While its profile shape exhibits canonical off-axis shoulders $\xi \approx 0.6$--$0.9$ matching~\citet{HusseinRoundJet}, near-axis levels rise steadily from $\approx 0.02$ at $z/l = 5$ to $\approx 0.09$ at $z/l = 17$ as the turbulence continues to equilibrate. The far-field profiles ($z/l = 13$--$17$) collapse onto one another in shape, confirming self-similarity, but as a group sit outboard of the canonical reference and their near-axis amplitude is still rising. The wake has therefore reached internal self-similarity without yet matching the canonical free-jet state, since turbulent energy deposited off-axis by the individual rotor shear layers needs additional distance to redistribute~\citep{TaddesseTwinJetLES}. 

The cross-stream turbulent normal stress $\langle v'v'\rangle/u_c^2$ reflects the multi-rotor origin in both peak location and amplitude (figure~\ref{fig:self_sim_fluc_vv}). Near-field stations ($z/l = 5$--$11$) exhibit off-axis maxima ($\xi \approx 0.5$--$0.7$) and an on-axis deficit. With downstream development, the on-axis stress in the diagonal cut rises from $0.033$ at $z/l = 5$ toward the canonical centerline-peaked form of~\citet{HusseinRoundJet}, which it approaches by $z/l\approx15$ and holds through the last station. This inward peak migration illustrates the wake's gradual relaxation from discrete rotor sources toward an axisymmetric state~\citep{OkamotoTurbulentJet, TaddesseTwinJetLES}, maintaining typical round-jet magnitudes of roughly half the streamwise component~\citep{WygnanskiPreservingJet}.

In contrast to the lagging turbulent normal stresses, the turbulent shear stress $\langle u'v'\rangle/u_c^2$ collapses cleanly (figure~\ref{fig:self_sim_fluc_uv}), reaching a peak of $\approx 0.023$ near $\xi \approx 0.8$--$0.9$ and matching~\citet{HusseinRoundJet} through the core. Since shear stress (\S\ref{subsec:Re_shear_stress}) is governed by the gradient transport of mean momentum, it equilibrates rapidly once the mean velocity achieves self-similarity.

Under uniform eddy-viscosity closure ($\langle u'v'\rangle = -\nu_T\,\partial\langle u\rangle/\partial r$), fitting the measured core shear stress ($0.3 \le \xi \le 1$) gives a normalized eddy viscosity $\hat{\nu}_T$ = $0.028$. This corresponds to a turbulent Reynolds number $\Rey_T = u_c r_{1/2}/\nu_T = 35.7$, matching the canonical round-jet value ($\Rey_T \approx 35$;~\citealp{PopeTurbulentFlows}). As an independent check, estimating $\hat{\nu}_T$ from the measured spreading rate yields $0.027$, consistent with the shear-stress fit. The closure model (figure~\ref{fig:self_sim_fluc_uv}, dashed) closely reproduces the measured core shear stress.

Together, these profiles reveal a clear separation of timescales. The mean flow merges into a single column by $z/l \approx 5$ and collapses by  $z/l \approx 13$, with Reynolds shear stress following in step. Conversely, turbulent normal stresses retain the four-rotor signature further downstream: the streamwise and cross-stream component reaches its canonical centerline-peaked form only by $z/l \approx 15$ in the diagonal cut. Comparing measurement planes highlights that while the mean downwash becomes axisymmetric, the underlying turbulence field does not. 

\section{Conclusions}
Planar PIV along two orthogonal cut planes characterized the transition of hovering quadrotor downwash from four discrete rotor sources to a single merged wake column ($z/l \le 17.5$). First- and second-order velocity moments relax to a self-similar form at different rates: the mean flow and Reynolds shear stress achieve canonical self-similarity by $z/l \approx 13$, whereas the turbulent normal stresses lag. The turbulent normal stresses begin approaching self-similarity near $z/l \approx 13$ and reach it by $z/l\approx 15$ in the diagonal cut, while the front-rotor cut retains its twin off-axis peaks to the end of the domain. Comparing the two cuts confirms that while the mean downwash becomes axisymmetric, the underlying turbulence retains strong structural memory of the four-rotor geometry. 

Given that a key feature of drone vehicle wakes is that they are generated by rotors, not a pressure source, there are two clear ways in which these measurements can be enhanced. Firstly, it would be important to measure the out-of-plane velocity fields. Secondly, there are likely to be distinct structures associated with each of the rotor flows - tip vortices, for example- and capturing the time- or phase-resolved structure related to the blade rotations (and perhaps exploring the effects of synchronization between adjacent rotors) is a natural next step to fully understand the behavior of these complex flows. 

Nevertheless, these findings have direct practical consequences for multi-UAV proximity flight ($\Rey_{D_{eff}} = 3\times10^4$). Downwash disturbances felt by a follower aircraft are driven primarily by this sheared, bimodal turbulence rather than the mean velocity deficit. Current disturbance-rejection methods for small UAVs were built for a different flow. Passive airframe features, such as bio-inspired leading edges, manage atmospheric gusts~\citep{DiLucaLowRe}; active schemes estimate aerodynamic loads from upstream multi-hole probes and pressure taps~\citep{FanFixedWing}. Both assume a broadly uniform gust field; however, a quadrotor downwash is not uniform. It is a localized, source-structured disturbance, so a follower likely needs control that knows where in the wake column it sits. Control frameworks and trajectory planners for close-formation UAV teams must therefore account for this decoupling between mean-flow axisymmetry and persistent turbulent anisotropy.

\section{Declaration of interests}
The authors report no conflict of interest 

\section{Funding \& Acknowledgments}
This work was supported by a Brown University Seed Award from the Office of the Vice President for Research. AK acknowledges support from NSF Graduate Research Fellowship (Award 2439559).

\bibliographystyle{jfm}
\bibliography{jfm}

\end{document}